\documentclass[12pt]{article}

\usepackage{graphicx}

\usepackage{ulem}

\usepackage[super]{natbib}

\usepackage{hyperref}

\usepackage{amssymb}
\usepackage{amsmath}
\usepackage{amsbsy}
\usepackage{amsfonts}
\usepackage[euler]{textgreek}

\usepackage{booktabs}
\usepackage{siunitx}
\usepackage{color}
\usepackage[dvipsnames]{xcolor}

\DeclareSIUnit{\calorie}{cal}

\usepackage{times}

\newenvironment{sciabstract}{%
\begin{quote} \bf}
{\end{quote}}

\title{Nanofluidic logic with mechano-ionic memristive switches}

\author
{Theo Emmerich$^{1,\dagger,\ast}$, Yunfei Teng$^{1,\dagger}$, Nathan Ronceray$^{1,\dagger,\ast}$,\\
 Edoardo Lopriore$^{2}$, Riccardo Chiesa$^{2}$, Andrey Chernev$^{1}$,\\ 
 Vasily Artemov$^{1}$, Massimiliano Di Ventra$^{3}$, Andras Kis$^{2}$,\\
and Aleksandra Radenovic$^{1,\ast}$\\
\\
\normalsize{$^{1}$Laboratory of Nanoscale Biology, Institute of Bioengineering}\\
\normalsize{Ecole Polytechnique Federale de Lausanne (EPFL), CH-1015 Lausanne, Switzerland}\\
\normalsize{$^{2}$Laboratory of Nanoscale Electronics and Structures}\\
 \normalsize{Institute of Electrical and Microengineering \& Institute of Materials Science and Engineering}\\
\normalsize{Ecole Polytechnique Federale de Lausanne (EPFL), CH-1015 Lausanne, Switzerland}\\
\normalsize{$^{3}$Department of Physics, University of California, San Diego, La Jolla, CA, 92093-0319, USA}\\
\normalsize{$^\ast$To whom correspondence should be addressed; E-mail:}\\ 
\normalsize{theo.emmerich@epfl.ch, nathan.ronceray@epfl.ch, aleksandra.radenovic@epfl.ch}\\
\normalsize{$^\dagger$These authors contributed equally}
}

\date{}

\begin{document} 


\baselineskip22pt


\maketitle 


\begin{sciabstract}
While most neuromorphic systems are based on nanoscale electronic devices, nature relies on ions for energy-efficient information processing. Therefore, finding memristive nanofluidic devices is a milestone toward realizing electrolytic computers mimicking the brain down to its basic principles of operation. Here, we present a nanofluidic device designed for circuit scale in-memory processing that combines single-digit nanometric confinement and large entrance asymmetry. Our fabrication process is scalable while the device operates at the second timescale with a conductance \textcolor{black}{ratio in the range 10-60.  \textit{In-operando} optical microscopy unveils the origin of memory, arising from the reversible formation of liquid blisters modulating the device conductance. The combination of features of these mechano-ionic memristive switches} permits assembling logic circuits composed of two interactive devices and an ohmic resistor. These results open the way to design multi-component ionic machinery, such as nanofluidic neural networks, and implementing brain-inspired ionic computations.

\end{sciabstract}
 
A fundamental difference between artificial computers and biological brains is the identity of their respective information carriers. Unlike computers relying on electrons and holes, living organisms deal with a wealth of different ions to process data\cite{voglis2006role}. Recent experiments have raised the prospect of  artificially reproducing this ionic computing, showing that nanofluidic channels filled with aqueous electrolytes can store information\cite{noy2023nanofluidic,xiong2023neuromorphic,robin2023long}. Nanofluidics aims at understanding and harnessing the peculiar properties of ionic and molecular transport under nanometric confinement\cite{kavokine2021fluids}. So far, this has been achieved in different geometries such as 0D-nanopores\cite{garaj2010graphene,feng2016single}, 1D-nanotubes\cite{li2023breakdown,tunuguntla2017enhanced,secchi2016massive}, or 2D-slits\cite{radha2016molecular,esfandiar2017size,gopinadhan2019complete,emmerich2022enhanced}. Using these platforms to emulate biological ionic transport\cite{agre2002aquaporin,gouaux2005principles,coetzee1999molecular,ajo2015crossing,lanyi2004bacteriorhodopsin,busath1981gramicidin} has been the subject of intense research\cite{tunuguntla2017enhanced,secchi2016massive,holt2006fast,xie2018fast,xiao2019artificial,zhang2013bioinspired,xu2021anomalous,davis2020pressure}.
A major goal for such biomimetic nanofluidics is the realization of computational operations, as nature relies on nanoscale ionic channels to perform information processing and storage at energy costs orders of magnitude lower than classical solid-state digital circuits\cite{noy2023nanofluidic,howard2012energy}. 

The discovery of a memristive effect in nanofluidic channels represents an important landmark\cite{noy2023nanofluidic,robin2023nanofluidics,hou2023learning,li2020synaptic,xie2022perspective}. A memristive device, sometimes loosely called a memristor\cite{pershin2022experimental}, is a passive two-terminal electrical component with a programmable conductivity that depends on its previous history of operation. Memristors can act as the artificial equivalent of biological synapses thanks to their ability to store information as a conductance value\cite{pershin2010experimental}\textcolor{black}{, thus enabling data storage and processing with a single device.} They can be used in electronic neural networks to adjust the connection strengths at nodes between crossbars, playing the role of the basic unit for brain-inspired, or neuromorphic, computing\cite{yao2020fully}. Unlike previous fluidic memristors\cite{sheng2017transporting,bu2019nanofluidic,zhang2019nanochannel}, micro-pipettes coated with polyimidazolium brushes and bidimensional slits revealed memory effects in simple electrolytes within the electrochemical window of water, specifically at voltages below 1.23 V\cite{xiong2023neuromorphic,robin2023long}. For both devices, this behavior is understood \textcolor{black}{using phenomenological modeling} by considering the combined effects of asymmetric entrances and slow interfacial diffusion, resulting in cycles of accumulation/depletion of ions inside the channel. Such a phenomenology recalls the induction of long-term potentiation through local changes in calcium ion concentration occurring in biological synapses\cite{evans2015calcium,kukushkin2017memory}. \textcolor{black}{Yet, direct experimental confirmations regarding the mechanisms of nanoscale fluidic memristors are still lacking, impeding knowledgeable nano-engineering efforts to improve the performances of these devices. Furthermore,} the large volume occupied by micro-pipettes and the complex fabrication procedure of 2D slits as well as the slow speed \textcolor{black}{or the} small hysteresis displayed by these devices hinders their use as synaptic components at the circuit scale. However, to enable practical computing applications like logical operations, pattern recognition, or image processing, it is essential to connect multiple memristive devices \cite{li2018analogue,chu2014neuromorphic,borghetti2010memristive}. This requirement has been identified as a crucial challenge in the field of nanofluidic computing\cite{noy2023nanofluidic,robin2023nanofluidics,bocquet2020nanofluidics,xiong2023fluidic} calling for a new generation of compact nanofluidic memristors that offer improved reliability, scalability, and performance. The goal of this work is to open the way for such brain-inspired computations with cooperative nanofluidic devices. We first present a device designed for circuit scale neuromorphic nanofluidics, called highly asymmetric channels (HACs), which cannot be classified as 0D, 1D or 2D. \textcolor{black}{By coupling electrokinetic measurements with optical observations we unveil unambiguously the origin of memory in HACs, arising from a combined mechano-ionic effect.}  We finally rely on HACs' combination of features to implement a nanofluidic logic operation by connecting two fluidic cells. 

To fabricate HACs (Figure 1.a and SI section 1), we start with silicon nitride (SiN) windows presenting a single circular aperture in their center (step 0). A thin discontinuous layer of palladium is then evaporated (step 1). Finally, a graphite crystal is deposited above the aperture (step 2). Since step 0 and step 1 are achieved at wafer scale, this current production approach is only limited by step 2 achieved at the single chip-scale. This step is however a simple dry-transfer of a large pristine van der Waals crystal not requiring high positioning accuracy. The main fabrication improvement lies in eliminating the intricate etching process necessary for creating 2D channels\cite{radha2016molecular,emmerich2022enhanced}. With further developments, the complete nanofabrication of HACs could be realized on the wafer scale. Nevertheless, the current process already enables us to produce HACs in batches of several tens of units by avoiding delicate fabrication steps at the single device level. This scalability is required to build nanofluidic circuits where several devices have to function simultaneously. In this study, we present results on a total of \textcolor{black}{32} devices referenced in Supplementary table 1 with corresponding experiments.

\begin{figure}[!h]
	\centering
	\includegraphics[width=1\linewidth]{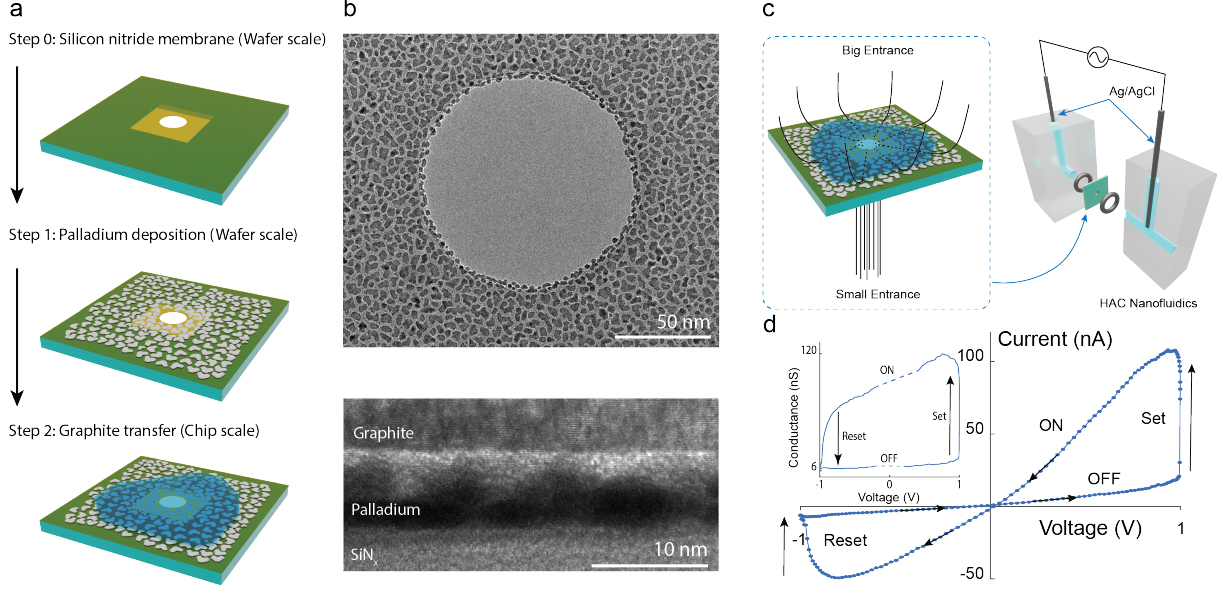}
	\caption{ \footnotesize	\textbf{Highly asymmetric channels.} \textbf{A,} Nanofabrication process flow. \textbf{Step 0:} The starting point are silicon chips (in green) with a silicon nitride membrane (in yellow) of around 20-by-20 microns, a thickness of 20 nm and an aperture with diameter of approximately 100 nm. \textbf{Step 1:} Palladium islands (in grey) evaporation. As a result of careful tuning of the deposition parameters, palladium islands form leaving space for ions to flow around them. \textbf{Step 2:} Dry transfer of a graphite crystal (in blue) with a lateral dimension of 20-50 micrometers. \textbf{B}, TEM images. Upper panel: top view of the SiN aperture after step 1.  Lower panel: a cross-sectional view of a completed HAC device. \textbf{C}, Device and setup. Left panel: sketch of a finished device with similar color than in \textbf{A}. Black lines indicates typical field lines/fluidic paths. The graphite crystal, aperture and islands are not at scale for clarity. Right panel: experimental setup for nanofluidic measurements. \textbf{D}, IV characteristics (Device A-50 mHz-1M KCl). The IV curve is composed of loops self-crossing at the origin, the signature of a memristive effect. Arrows indicates the direction of the sweep. The dots show individual data points highlighting the abrupt switching and quasi-discrete conductance states. Inset: GV curve extracted from \textbf{D}. The conductance is defined as the instantaneous current/voltage ratio, $G(t)=i(t)/v(t)$.  Here, the conductance is between 6 and 120 nS yielding a conductance ratio of 20.}
\end{figure}

TEM characterization provides information regarding the geometry of HACs (Figure 1.b and Supplementary Figures 2 and 3). Pd islands have a characteristic lateral dimension of 5-10 nm, a spacing and a height of a few nanometers. Ions and water molecules experience single-digit nanometric confinement by flowing around palladium islands, converging to or diverging from the SiN aperture depending on the sign of the applied potential. HACs present a large asymmetry between their two entrances as highlighted in Figure 1.c.  The first entrance, through the membrane aperture, has a lower area $A_{l}=\pi Dh$ with where $h$ is the islands height and $D$ is the aperture diameter. The second entrance through the crystal edge has a larger area $A_{h}\simeq\pi Lh$, where $L$ is the crystal characteristic dimension comprised between 20 \textmu m and 50  \textmu m. Thus, the ratio of entrance areas $\frac{L}{D}$ is on the order of several hundred, depending on the size of the top layer crystal. We expect that this ratio will promote ionic accumulation when counter-ions enter the channels through the crystal edge and depletion when they enter through the aperture according to the theoretical framework presented by Robin \textit{et al}.\cite{robin2023long}.

To perform nanofluidic measurements, HACs are placed into a fluidic cell separating two reservoirs filled with potassium chloride aqueous solution \textcolor{black}{at pH 5.5} (Figure 1.c). Upon the application of a sinusoidal potential, HACs exhibit a clear bipolar memristive signature\cite{robin2023long} (the conductance increases at positive voltage and decreases at negative ones with memory retention at low voltage as shown in Figure 1.d) favorable for computing operations with programming voltage pulses. At 1M KCl, they operate at frequencies in the 30-300 mHz range with a conductance ratio \textcolor{black}{between 9 and 60 depending on the device (Figure 1.d and Supplementary table 1)}. The memory effect in HACs is therefore typically one order of magnitude stronger and two orders of magnitude faster compared to slits exhibiting similar bipolar dynamics \textcolor{black} {while presenting a larger hysteresis than micro-pipettes}\cite{xiong2023neuromorphic,robin2023long}. Our improved performance allow completely setting/resetting HACs with 2 seconds voltage pulses (Supplementary Figure 6). \textcolor{black}{We additionally verified that HACs can be set to intermediate levels by applying short voltage pulses, showing the possibility of implementing neuromorphic functions such as spike-timing dependent plasticity (Supplementary Figure 7).} Besides exhibiting a combination of fast speed and large conductance ratio, the response of HACs also qualitatively differs from previous devices by exhibiting delayed switching\cite{wang2010delayed}. When the applied voltage becomes positive, the response is initially linear until a given threshold is reached where the conductance dramatically increases (Figure 1.d). This abrupt transition from the OFF state to the ON state justify referring to this phenomenon as \textit{switching}. Such  dynamics is reminiscent of solid-state electrochemical metallization memory cells (ECM)\cite{valov2011electrochemical} and have not been reported in nanofluidics. The switching threshold \textcolor{black}{(ON/OFF behavior)} enables reading the memory state without disturbing the programmed state and is beneficial for several neuromorphic computing applications such as bio-realistic Hebbian learning or conditional logic\cite{borghetti2010memristive,boyn2017learning}.  These initial findings demonstrate the successful fabrication of scalable nanofluidic switches that exhibit unparalleled performance and operate effectively using simple monovalent salt solutions while staying within the electrochemical window of water. 

The \textcolor{black}{nature} of the threshold behavior is studied by tuning the applied sinusoidal bias. The voltage at which the switching occurs is dependent on the bias amplitude (Figure 2.a and Supplementary Figure 8 with two additional devices). Consequently, switching is not \textcolor{black}{triggered at a specific voltage threshold value, regardless of the chosen voltage waveform (Supplementary Figure 9)}. Current traces recorded at different frequencies highlight the switching nature of ionic transport in HACs (Figure 2.b and Supplementary Figure 11), occurring at timescale faster than 100 ms. The delay time before switching, $\tau$ increases with decreasing frequency of the applied bias signal. Regardless of the applied frequency, the switching occurs when a given amount of charge $Q$ has flown out of the device (Figure 2.b-c). We recover such conservation of the charge threshold with three additional devices, as shown in  Supplementary Figure 11. \textcolor{black}{The complete IV cycles at various frequencies, leading to linearity at higher frequencies as expected for a memristor, are displayed in Supplementary Figure 10  for four devices. We also show in  this supplementary Figure frequency sweeps down to 1 mHz showing a diode-like behavior corresponding to the low-frequency limit of a charge-threshold bipolar memristor.}

\begin{figure}[!h]
	\centering
	\includegraphics[width=1\linewidth]{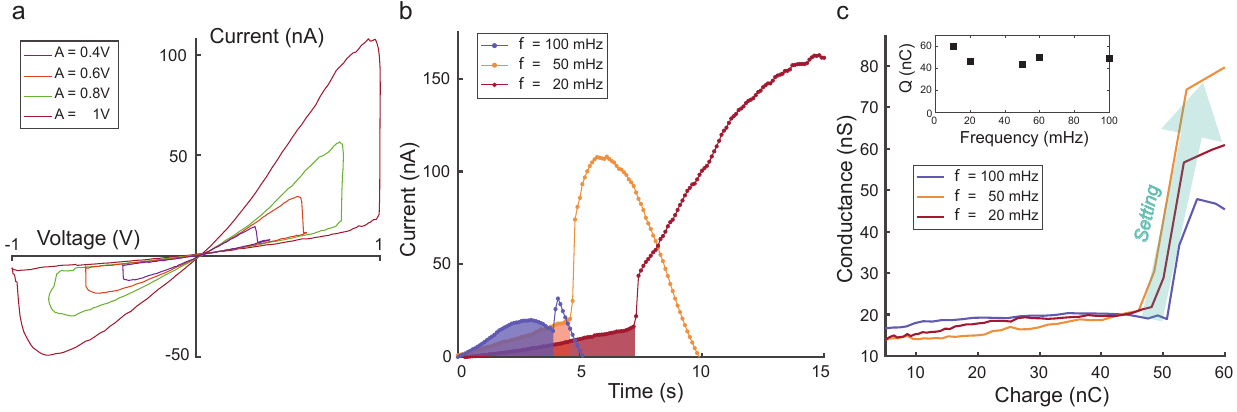}
	\caption{ \footnotesize\textbf{\textcolor{black}{Nature} of the switching threshold (Device A-1M KCl).} \textbf{A,} IV characteristics for different applied sinusoidal potential amplitudes. The applied bias frequency is 50 mHz. \textbf{B,} Device switching at different frequencies of the applied positive sinusoidal bias voltage with an amplitude of 1V. The dots are individual data points. The shaded area represents the charge threshold $Q=\int_{0}^{\tau}i(t)dt$, where  $i(t)$ is the measured current. This quantity is conserved across frequencies. \textbf{C,} Dependence of the conductance on the cumulative amount of charge, $q(t)=\int_{0}^{t}i(t)dt$ flowing out of the device, extracted from \textbf{B}. The conductance is the instantaneous current/voltage ratio. The conductance abruptly increases when the charge threshold is reached. Here the charge threshold is approximately equal to 50 nC. Inset: Frequency dependence of the amount of charge $Q$ flowing out of the device before reaching the threshold.}
\end{figure}

 \textcolor{black}{To understand the origin of such memristive dynamics we observe HACs \textit{in-operando} through wide-field light reflection imaging as illustrated in Figure 3.a. Such direct observation of nanofluidic  processes, and in particular memory,  has been identified as a major challenge to be reached for the field\cite{robin2023nanofluidics,bocquet2020nanofluidics}. Briefly, the device chip is mounted in a custom fluidic cell similar to Figure 1.c, but with a top reservoir large enough to accommodate a water-dipping objective. The reflected light intensity from the multi-layer structure of the device is set by thin film interference, which was previously shown to enable monitoring deformations of 2D devices at the 10 nm scale\cite{Ronceray2023}. See SI section 4.1 for details regarding in our \textit{in-operando} optical setup}
 
 \textcolor{black}{Our combined electrokinetic and optical results are presented in Figure 3.b. Concomitantly with the IV curve acquisition, we observe the emergence of  interference fringes at positive voltage and their disappearance at negative ones. These patterns arise from the formation of a (sub-)micrometric liquid blister between the Pd/SiN surface and the graphitic one. Since monochromatic light is used for imaging (λ=635 nm), a contrast inversion corresponds to a blister height difference of half a wavelength. Indeed, the blister modulates ion conduction dramatically by suppressing HACs' resistance at its bottleneck, close the SiN aperture, explaining the larger conductance at positive voltage and overall bi-stability. }
 
 \textcolor{black}{For the devices in Supplementary Movies 1 and 2, the switching occurs when the blister edge moves across the SiN aperture (Figure 3.b-d). In this mechanism, the conductance state of the device correlates perfectly with reflected intensity signal collected from the pore region (integrated in a 3x3um window around the pore), as shown in Figure. 3.c. As observed in Figure. 3.b, when a small off-centered blister is present, the conductance is low (OFF-state, step 0), but after the threshold corresponding to the enlarged blister crossing the pore (step1) the conductance reaches a higher value (ON-state, step 2). Within the simplified radial model introduced in SI Section 5, the conductance ratio can be expressed as: $G_\text{ON}/G_\text{OFF} = ln(r_\text{out}/r_\text{in})/ln(r_\text{out}/r_\text{blister})$. With $r_\text{in}$=50 nm, $r_\text{out}$=50 \textmu m, $r_\text{blister}$=30 \textmu m, this formula predicts the right order of magnitude of observed conductance ratios.}

 \textcolor{black}{ In other cases, the blister forms directly above the hole and we thus attribute the switching to a sudden increase of blister volume taking place outside the SiN window (Supplementary Movies 3-4). This dichotomy points to the importance of optimizing HACs in order to control the blister motion and therefore the performances of the memristor. This will enable the building of neuromorphic nanofluidic circuits with a large number of components.  All videos containing matching optical signals and IV plots can be found in Supplementary materials.}

  \textcolor{black}{We attribute the formation of the blister to electrostatic forces surpassing the van der Waals adhesion forces between the palladium islands and the graphite crystal as illustrated in Figure 3.e. Indeed, while the large surface charge of graphite in 1M KCl ensures a net cation excess $\rho_+$ in the device at rest, the high channel asymmetry enables focusing this net charge when the electric field points inwards (positive voltages). The obtained out-of-equilibrium net charge density builds up large electrostatic pressures scaling as $P_{+}=Kr_{Pd}^2\rho_+^2/ \epsilon_0\epsilon_r$, where K is a geometric prefactor (derived in SI section X as $K={ 2\pi\over15}$ for the simplified case of a spherical interstitial space)  and $r_{Pd}$ is the palladium island typical size. When cations are focused enough, it becomes energetically favorable for the system to relax through blister formation when the net charge density exceeds its critical value defined by $\rho_+^*=\sqrt{\frac{\Gamma\epsilon_0\epsilon_r}{Kr_{p}^{3}}}$ where \textGamma \hspace{0.1cm} is the adhesion energy of graphite on palladium. For $r_{Pd}$=5nm, $\Gamma$=20mJ/m$^2$ and an outer surface charge of  0.1C/m$^2$, we obtain that a 2-fold out-of-equilibrium net charge density increase is enough to trigger blister formation. Our analytical calculations reveal that counter-ions can accumulate due to their radially converging trajectory inside the device, overcoming diffusion. The charge focusing is found to be rather broad for the voltages applied, which explains that the blister does not systematically form at the center. Additionally, it can take different shapes depending on the device geometry and pre-strain, resulting in the observed device-to-device variability. The application of negative voltage  flushes excess cations out, enabling the blister to relax to lower its energy through adhesion. Although we rationalized the blister filling and emptying through the effects of charging and adhesion only, electro-osmotic flow is likely to play a role in the observed water motion.  Overall, HACs are ruled by the ratio $\Sigma^2\over\Gamma$, which should be large enough to achieve significant blister dynamics and therefore memory. Further details and aforementioned analytical calculations on the mechanism are provided in Supplementary section 5.}
 
  \begin{figure}[!h]
	\centering
	\includegraphics[width=1\linewidth]{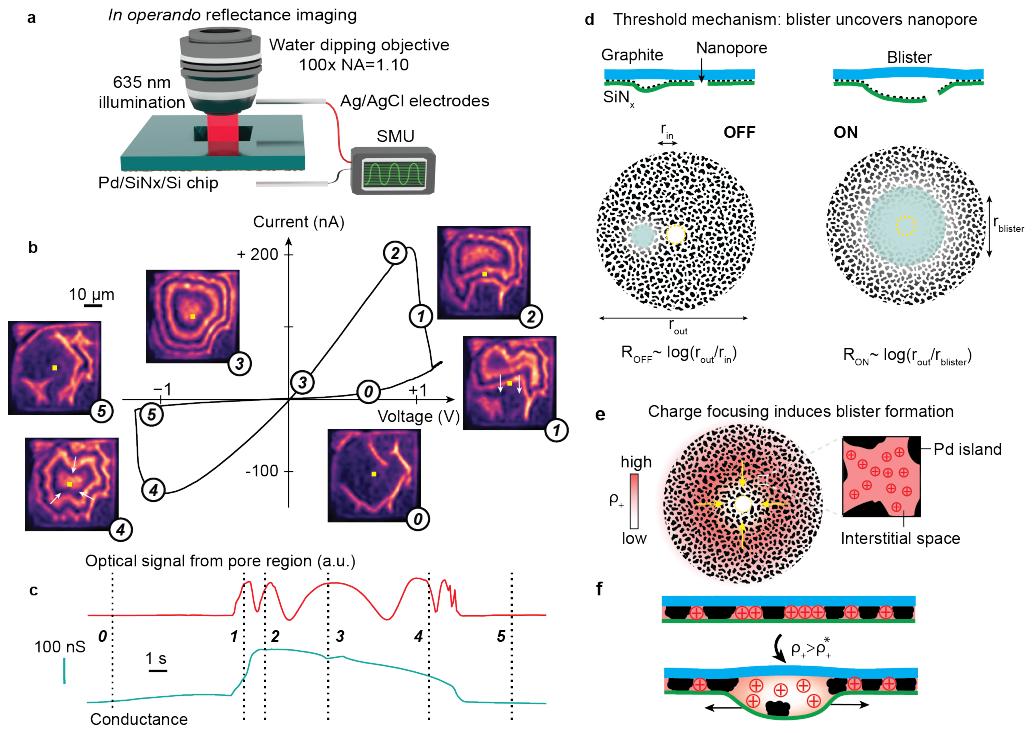}
	\caption{ \footnotesize\textbf{In-operando microscopy reveals the memory mechanism. a},  Setup schematic for correlative electrokinetic and optical measurements. The reflected signal from the chip backside is collected with a water dipping objective while measuring ionic current. \textbf{b}, IV characteristic (Device 2, 1V, 30 mHz) with optical images of the SiNx window at different time points marked by white dots and numbers. White arrows illustrate blister motion. \textbf{c}, Optical signal from the pore region denoted by the yellow square in the images in \textbf{b} and its correlation with the device conductance during the IV acquisition. \textbf{d}, Threshold mechanism observed in \textbf{c}: a liquid-filled blister forms at positive bias (branch 0-1 in \textbf{b}) between the graphite and Pd/SiNx walls, and the threshold occurs when the blister front crosses the nanopore (1 in \textbf{b}). \textbf{e}, The blister formation occurs through the excess positive charge focusing in the radially converging channel network, yielding an increased cation density in the interstitial spaces between palladium islands. \textbf{f}, When this density exceeds a critical value, a blister forms to relax the electrostatic energy. Nanoscale blisters as pictured in \textbf{e} results in micron-scale blisters under sustained positive biases, yielding features spanning the whole SiNx window as shown in \textbf{b}.}
\end{figure}

 We verify our key assumptions with a set of control experiments. \textcolor{black}{Using programming pulses, we notice that HACs remain in a high conductance state after setting, similar to synaptic long-term potentiation as the blister requires a negative voltage to vanish (Supplementary Figures 12 and 21), confirming that conductivity is directly correlated to the blister state.} Pressure-driven streaming currents  \textcolor{black}{as well as osmotic currents induced by concentration gradient} show that HACs exhibit K$^+$ ionic selectivity arising from a net negative surface charge (Supplementary Figure 16). The conductance of HACs drops by two orders of magnitude between 1M and 1mM KCl and the memristive effect can only be observed for concentrations above 100 mM (Supplementary Figure 14). This points to the importance of the surface charge which increases, \textcolor{black}{for mechanically exfoliated pristine graphite as used in this work}, with the salt concentration to reach an absolute value above 0.1 C/m$^2$ at 1M\cite{emmerich2022enhanced}. \textcolor{black}{Molecular dynamics simulations also showed that the interface of pristine graphite and hexagonal boron nitride (h-BN) get electrified in water\cite{grosjean2019versatile}}. While the effect can still be observed with a h-BN cover layer, this is not the case with mica (supplementary Figure 15). \textcolor{black}{We also performed \textit{in-operando} optical measurements with mica controls directly showing considerably reduced blister formation occurring at negative voltages, as opposed to the large deformations observed for graphite at positive voltages only, apparently pointing to a lower $\Sigma^2\over\Gamma$ (Supplementary Figure 22). The emergence of the effect with graphite cap at high salt concentration (equivalent to high surface charge) and the highly reduced blister dynamics with mica indicates that the memory effect requires a high charge of the top layer interface. For this reasons, as well as because their spacing is larger than twice the Debye length at 1M (0.3nm), we neglect the contribution of palladium islands on the ionic transport and consider them as neutral spacers. They however play a role on the memristive effect through the adhesion energy, \textGamma.}
 
  \textcolor{black}{The origin of memory in HACs is thus related to reversible mechanical deformation induced by the presence of excess counter-ions, explaining the observed charge threshold for switching.} The existence of charged zones resulting in deviations from local electroneutrality has already been used to describe out-of-the-equilibrium and non-linear ion transport phenomena at the nanoscale\cite{rubinstein2010extended,rubinstein2010dynamics,jubin2018dramatic}.  \textcolor{black}{However, the large electrostatic forces that can build up in nanoscale systems breaking electroneutrality were never reported to trigger mechanical deformations at larger scale.} HACs represent the optimal geometry to favor such effects by combining large asymmetry with single-digit confinement in the presence of a highly charged interface, resulting in dramatic \textcolor{black}{mechano-ionic} memristive dynamics.

Having established scalable nanofluidic device fabrication and \textcolor{black}{obtained direct experimental proof regarding} the underlying switching mechanism, we can utilize this knowledge to harness the potential of HACs in the realm of ionic computing.  Building a logic circuit with two aqueous memristors influencing each other represents a paradigm shift for nanofluidics where devices have until now been measured independently. We can now wire them for neuromorphic computing applications by taking advantage of the performances and reliability of HACs. This confers a new purpose to nanofluidic devices, in addition to their role as a technological platform for uncovering fundamentals of molecular transport at the smallest scales. Using two parallel HACs connected to a resistor as shown in Figure 4.a, we implement the material implication (IMP) logic gate as demonstrated by Borghetti \textit{et al}.\cite{borghetti2010memristive} with solid-state memristors. The first HAC is called the P-switch and its conductance state remains unchanged during the logic operation. The second one is the Q-switch. Switching it from the OFF state (low conductance, defined as 0) to the ON state (high conductance, defined as 1) is only possible when P is in a low conductance state. Such conditional switching enables the implementation of the the first two rows of the IMP truth table (Figure 4.b), being the non-trivial cases. The two last rows are vacuous truth (or trivial cases) as Q is already switched-on at the start of the logic operation. In this demonstration of ionic computing, both nanofluidic memristors interact through the resistor to realize a conditional logic operation. The IMP gate represents a milestone for nanofluidic memristive action, as it can be used to derive any other classical logic gate commonly employed in digital computing\cite{borghetti2010memristive}.

\begin{figure}[!h]
	\centering
	\includegraphics[width=1\linewidth]{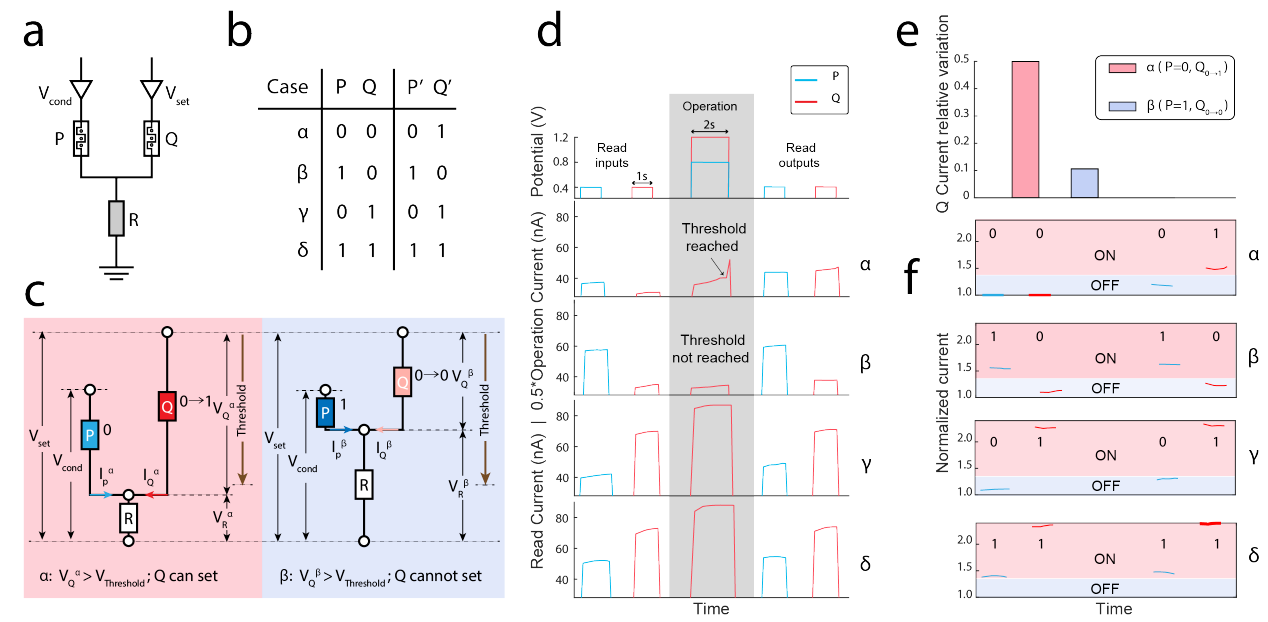}
	\caption{ \footnotesize \textbf{Nanofluidic logic.} \textbf{A}, Circuit schematic. Two HACs are connected in parallel with a variable resistor set to 6 MOhm. The working electrode of each device are connected to the channels of a source-measurement unit. The shared ground electrode of the two cells is connected to the resistor. \textbf{B,} IMP truth table. The first two columns (P and Q) represent input states, and the neighboring P' and Q' columns are the corresponding outputs of the IMP gate. A Greek letter is allocated to each one of the logic cases. \textbf{C,} Illustration of the working principle of conditional switching implementing the non-trivial cases of the IMP truth table. Left panel represents $\alpha$ case and right panel represents $\beta$ case.   The effective voltage $V_\text Q$ applied to the Q-switch is sufficient to reach the charge threshold of the Q device within the pulse duration when P is in the 0 state ($\alpha$ case) and insufficient when P is in the 1 state ($\beta$ case). \textbf{D,} Applied voltage (first row) and measured ionic traces on both Q and P lines (Devices B and C, 1M KCl). The measured pulses before the operation give the IMP table inputs and the ones after the operation provide the outputs. The Q-line current during operation (with grey background) reaches the threshold in the $\alpha$ case, indicated by an abrupt increase. \textbf{E,} Read current relative variations of the Q-line for the $\alpha$ case (rose background, P=0) and $\beta$ case, (blue background, P=1). \textbf{F,} Read currents normalized by their respective minimum value in the $\alpha$ case both devices. The transition between the range corresponding to state 0 (blue background) and state 1 (rose background) occurs when the conductance of the corresponding device increases by at least 40\% relative to its minimum value. }
	
\end{figure}

As HACs can be set and reset at least ten times (Supplementary Figure 17), they are suitable for building logic circuits. \textcolor{black}{We also checked the endurance of HACs under sinusoidal operation (Supplementary Figures 18 and 23). We observe through electrokinetic measurement as well as optically perfect stability up to 17 cycles. We could also measure 300 cycles on another device, showing HACs robustness and setting an upper-bound for nanofluidic memristors.} Individual HACs can still operate when connected in series with the aforementioned resistor, even though  this reduces the effective conductance ratio (Supplementary Figure 26). To connect the switches, we use a single electrode with two chlorinated tips grounded through a variable resistor supplementary Figure 26). Both P and Q switches are first placed into their targeted initial conductance state by applying a  2 s voltage pulse of high amplitude (+ 1 V for state 1 and -1 V for state 0). Input states of the memristors are then probed by applying read pulses to the working electrodes of each switch. The IMP gate is operated by applying voltage pulses simultaneously to each switch (Figure 4.c). In particular, a conditional voltage pulse of amplitude $V_\text{cond}$, which is not high enough to reach the switching threshold, is applied to the working electrode of the P switch while a larger voltage pulse, $V_\text{set}$ is applied to the working electrode of the Q switch. When the P switch is in the OFF state, the current $I_\text{P}^{\alpha}$ flowing across it is of a lower value ($\alpha$ case shown in the left panel of Figure 4.c). Thus the voltage drop across the resistor $V_\text{R}^\alpha=R(I_\text{P}^\alpha+I_\text{Q}^\alpha)$ is sufficiently low, resulting in a large enough voltage drop $V_\text{Q}^{\alpha}$ across  Q, to reach the charge threshold required to switch Q. Inversely, when P is in the ON state ($\beta$ case shown in the right panel of Figure 4.c), the $I_\text{P}^\beta$ and $V_\text{R}^{\beta}$ are larger, making the voltage drop across Q ($V_\text{Q}^{\beta}$) insufficient for switching Q. This is why a switching threshold is essential for a successful implementation of IMP logic as illustrated in fig. S12 where we observe in single device experiments that setting HACs is only possible with 2 s programming pulses having an amplitude high enough for the ESC to reach the SiN aperture. The value of the resistor R is chosen such that the potential drop given by the low conductance state current of P is low enough to enable the pulsed setting of the Q switch. On the other hand, the value of R needs to provide a sufficient potential drop at the high-conductance state of P to avoid the setting of Q. R is therefore chosen such that it has a value between the devices resistances in the set and conditional states of the pulsed operation, $R_\text{SET}<R<R_\text{COND}$\cite{borghetti2010memristive}. Finally, the output states are recorded similarly to the input states. 

Figure 4.d displays the experimental results obtained for the IMP gate implementation with memristive HACs.  The Q-line current trace during the IMP gate operation reaches the switching threshold when P is set to 0 ($\alpha$ case) but not when P is set to 1 ($\beta$ case). The conductance of Q increases by 50\% in the $\alpha$ case and only 10\% in the $\beta$ case (Figure 4.e). We are thus able to condition the switching of a nanofluidic memristor by the conductance state of another device. We recover the IMP truth table by normalizing the read pulses of each memristor by their current minimum, while defining \textcolor{black} {\textit {a posteriori} an arbitrary} borderline delimiting ranges corresponding to states 1 and 0 (Figure 4.f). \textcolor{black} {Its existence serves as  a proof of concept, demonstrating  the potential to construct logic circuits using nanofluidic memristors as building blocks.}

For robustness, we carry out another conditional switching experiment with two HACs and show switching  with a HAC, as Q-switch and a variable resistor, as P-switch that emulates another HAC (Supplementary Figures 28 and 29). We have also performed numerical simulations of an equivalent electrical circuit consisting of  two charge-threshold memristive switches with \textcolor{black}{typical} conductance values and ratio as in our \textcolor{black} {pulse programming} experiments. 
\textcolor{black} {Qualitatively, simulation} reproduces experimental results, \textcolor{black}{showing that IMP logic is indeed achievable with HACs} (Supplementary Figure 30).  

Logic operations with HACs are made possible by the presence of a charge switching-threshold, a fast speed enabling operations with 2 s programming pulses, as well as a large conductance ratio compensating the performance loss caused by the presence of the resistor. The combination of these features is exclusive to HACs, making logic gating unattainable with previously reported nanofluidic memristors.

HACs are scalable and compact nanofluidic memristors that can be set or reset within a few seconds with a conductance ratio reaching \textcolor{black}{sixty}. \textcolor{black} {\textit{In-operando} optical observations show that their large entrance asymmetry and single-digit confinement result in a switching behavior related to the reversible formation of liquid blisters. This suggests several parameters that can be tuned for further optimization. Choosing the membrane size and stiffness, the surface charge and adhesion properties of materials as well as introducing guidelines in order to control the blister position and dynamics will help increase HACs performances and reliability as well as reduce the device-to-device variation. We unveil a nano-engineering playground aiming at creating better performing mechano-ionic memristive switches displaying programmable and dynamic geometries in order to regulate neuromorphic nanofluidic circuits.} HAC's strongly non-linear dynamics allow us to implement a Boolean operation with two interacting devices, thereby achieving the fundamental building block for future aqueous computing machines with increasing complexity. We demonstrate that geometrical effects at the nanoscale unlock the design of liquid hardware reminiscent of biological neural circuits. While \textcolor{black}{our HACs} at this point cannot match the performance of solid-state  Pt/TiO$_2$/Pt memristive switches, operating at the microsecond timescale\cite{borghetti2010memristive}, further efforts like  \textcolor{black} {optimizing their design} and connecting them with water channels in order to fabricate fully liquid circuits should result in improvements. By taking inspiration from electronic crossbar arrays as well as brains of living organisms, our results indicate a path toward nanofluidic neural networks.

\bibliographystyle{unsrt}

\pagebreak
\setlength\parindent{0pt}

\textbf{Funding:} T.E. N.R. and A.R acknowledge support from  European Union’s H2020 Framework Programme/ERC Advanced  Grant agreement number 101020445.
\textcolor{black}{T.E. acknowledge support from Swiss National Science Foundation grant No. TMPFP-2217134}.  E.L., R.C. and A.K. acknowledge funding from the European Union’s Horizon 2020 research and innovation program under grant agreements No 894369 (Marie Curie Sklodowska ITN network “2-Exciting”) and No. 881603 (Graphene Flagship Core 3 Phase) as well as the Swiss National Science Foundation (grants no. 157739 and 205114).M.D. is supported by the NSF grant No. ECCS-2229880.\\
\textbf{Author contributions:}
A.R. and T.E. conceived the project and designed the experiments; A.R. supervised the Laboratory of Nanoscale Biology team and A.K. supervised the Laboratory of Nanoscale Electronics and Structures team; \textcolor{black}{N.R. built the microscope and T.E built the fluidic cell for \textit{in-operando} optical measurements.} E.L. and R.C. built the ionic logic setup with inputs from Y.T. and T.E.; T.E., Y.T. \textcolor{black}{and N.R.}  performed the measurements with inputs from E.L. and R.C.; Y.T. made the silicon nitride windows and performed palladium deposition, T.E. transferred the 2D crystals; A.C. did the TEM cross-sectional image; N.R. developed the theoretical modeling with inputs from T.E. and V.A.; T.E \textcolor{black}{and N.R.} analyzed the experimental data.; Y.T. performed the numerical simulations with inputs from A.R.; T.E. wrote the manuscript with inputs from Y.T. and N.R.; all authors contributed to editing of the manuscript.\\
\textbf{Acknowledgments:} We  would  like  to  thank  the  EPFL  Center  of  MicroNanoTechnology  (CMi)  for their  help  with  chip  fabrication  and  the  EPFL  Center  for  Electron  Microscopy  (CIME)  for  access  to electron  microscopes, in particular Luice Navratilova an Victor Boureau, for assistance with TEM cross-sectional imaging. We thank Elison Matioli and Guilherme Migliato Marega for fruitful discussions.\\
\textbf{Competing interest:} none.\\
\textbf{Data and materials availability:} All data required to convey scientific findings described in the paper are present in the main text
or the SI.

\pagebreak
\begin{center}
\Large{\bf Methods}
\end{center}

\textbf{Device fabrication.} To fabricate HACs, we start with homemade silicon nitride membranes having a single hole of approximately 100nm wide at their center (Figure 1.a, step 0 and SI Figure 1). Using a electron-beam evaporation, we then deposit a discontinuous layer of palladium that self-organizes into clusters or "islands" leaving space for ions to flow around them (Figure 1.a, step 1 and SI Figure 2). Finally, we transfer a bidimensional crystal above the hole to close the system using a droplet-shaped Polydimethylsiloxane stamp covered with Polypropylene carbonate (Figure 1.a, step 2 and SI Figure 3). Here, we mostly used graphite but also tested the effect of using hexagonal boron nitride and mica as top layer material. Further details are given in SI section 1.\\

\textbf{Nanofluidic measurements.}  Ag/AgCl electrodes are used to apply the potential and measure the resulting current. Data acquisition setup for single device measurements consists of a \textit{FEMTO} amplifier (DLPCA-200) and a digital/analog converter (\textit{NI} 63 series). The measurements are sampled at 100 kHz with an acquisition time of 100 ms. We used KCl, CaCl$_2$ and AlCl$_3$ solutions with concentrations ranging from 1mM to 1M. For logic experiments with two channels, the working electrodes of each cell are connected to a dual channel Keithley source-measurement unit (2636B model) using a sampling rate of 10 Hz.

\end{document}


\thispagestyle{empty}

\singlespacing


\baselineskip22pt


\maketitle 
\setcounter{page}{1}

\tableofcontents{}

\clearpage

\section{Nanofabrication}
\subsection{Step 0: Wafer-scale fabrication of silicon nitride aperture}

Silicon nitride (SiN) windows, as the basis and the starting point of our HAC nanofluidic devices, are fabricated in a state of art cleanroom facility with a technical route of reactive ion etching (Supplementary Figure 1.a)\cite{thakur2020wafer,verschueren2018lithography}. Here, we start from a 4-inch double side polished silicon wafer with a 60 nm SiO$_2$ and a 20 nm low stress silicon nitride at both sides from Center of MicroNanoTechnology in EPFL. The first step is the back-side patterning. Direct-laser pattering and reactive ion etching will result in a back-side design on the silicon nitride and silicon oxide layer (Supplementary Figure 1.b), which includes working membrane positions, contrast markers and dicing lines (supplementary Figure 1.d). Then, the KOH wet etching will release the freestanding silicon nitride membrane according to the pattern from the previous step. In the end, we pattern the front side design and silicon nitride apertures with electron beam lithography and reactive ion etching. With several cleaning procedures and physical dicing, we can produce these silicon nitride nanopore chips (Supplementary Figure 1.c) in a wafer-scale fabrication process flow.

\begin{figure}[!h]
	\centering
	\includegraphics[width=1\linewidth]{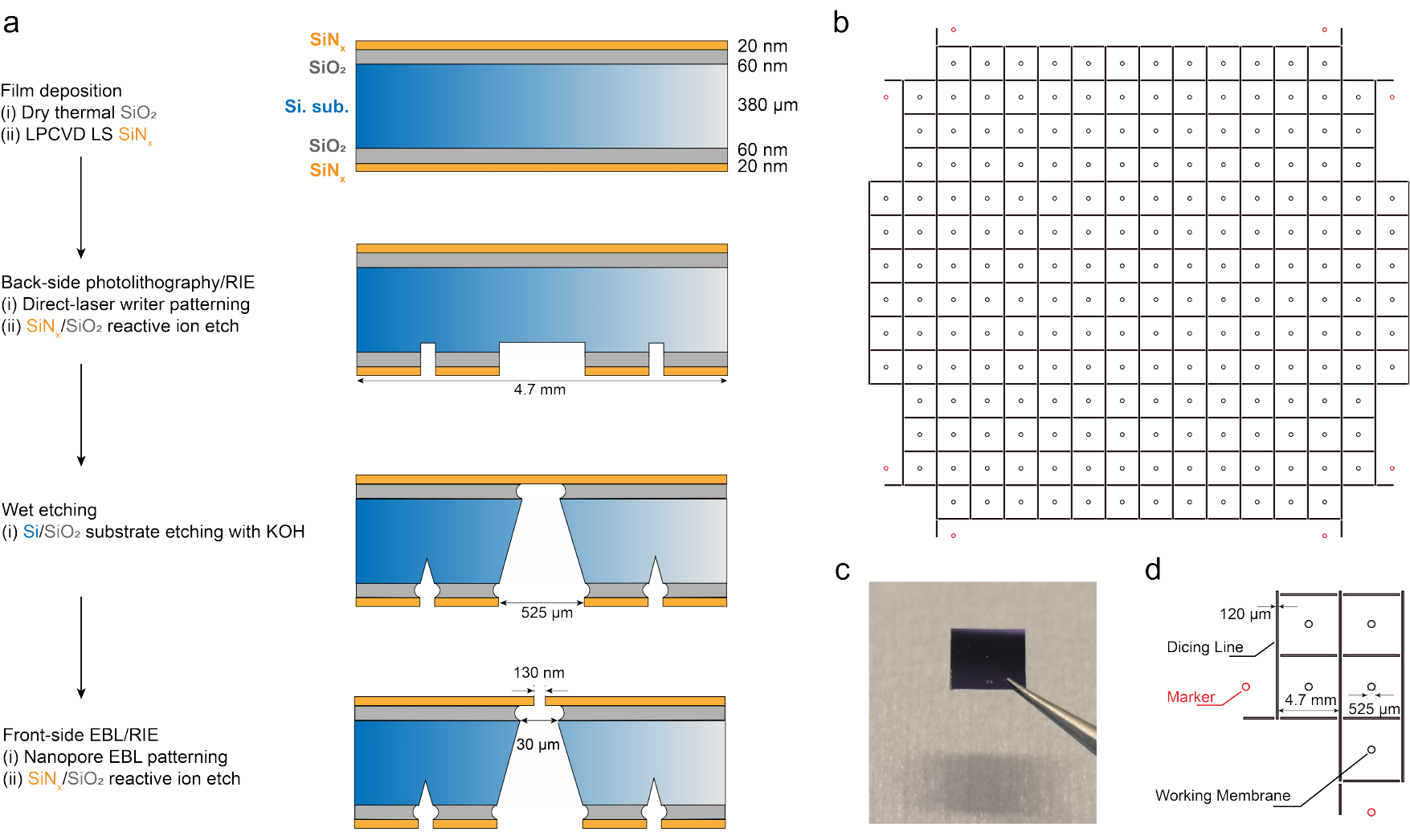}
	\caption{ \footnotesize	\textbf{Wafer-scale fabrication of silicon nitride aperture. a,} A fabrication process flow based on the reactive ion etching to realize pattern transfer from PMMA photoresist to silicon nitride membrane. \textbf{b,} Pattern design on the backside of the wafer. \textbf{c}, Optical image of a single chip. \textbf{d,} The detailed dimensions of pattern design in \textbf{b}.}
\end{figure}
\pagebreak
\subsection{Step 1:  Palladium deposition and characterization}
Electron beam deposition is usually used to introduce metallic materials with a well controllable deposition rate. Due to the mechanism of electron beam vapor deposition, the landing of metal atoms on the surface of the target can also be taken as a condensation of atomic vapor. Thus, for depositing the metal that is not very adhesive on the silicon nitride surface, the usual process consists of depositing first an adhesive layer such as Ti or Cr. However, the direct deposition will cause a non-continuous and non-uniform layer of metal on the silicon nitride surface (supplementary Figure 2). \cite{kiefer2010transition}

\begin{figure}[!h]
	\centering
	\includegraphics[width=1\linewidth]{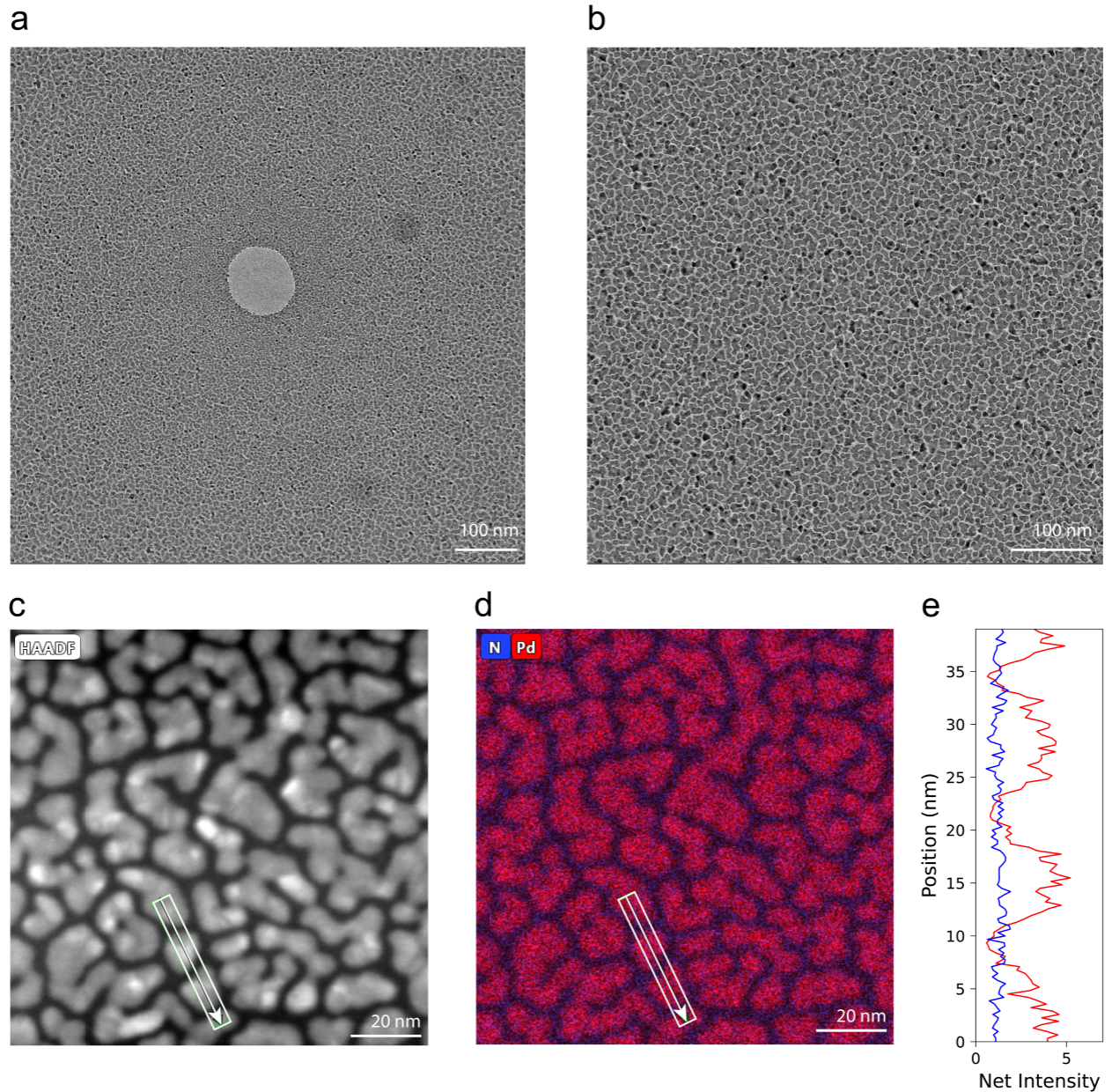}
	\caption{ \footnotesize	\textbf{Transmitted electron microcopy analysis of non-continuous and non-uniform palladium layer on free-standing silicon nitride membrane} This non-continuous and non-uniform morphology is consistent on the surface of silicon nitride membrane around the aperture \textbf{(a)} or far from the aperture\textbf{(b)}. With the scanning transmitted electron microscope, the high angle annular dark field (HAADF) image \textbf{(c)} and EDX mapping \textbf{(d)} are provided to support the non-uniform distribution of Pd on the uniform substrate of silicon nitride. \textbf{(e)} is an extracted line analysis of Pd and N net intensity.}
\end{figure}

For instance, here is the result that we deposited a very thin layer (~2 nm) of palladium directly on the surface of silicon nitride aperture. Due to the condensation effect during metallic layer formation and the poor adhesive property of Pd on silicon-based surface, a non-continuous layer of Pd is formed with isolated island geometry (Supplementary Figure 2.a). This non-uniform morphology of Pd layer is due to the adhesive relation between metal and targeted substrate. Therefore, this effect is consistent on the surface of silicon nitride membrane, the specific geometric feature, Pd island, could be observed both around (Supplementary Figure 2.a) and far from the silicon nitride aperture (Supplementary Figure 2.b). Furthermore, the EDX mapping (Supplementary Figure 2.d) is also provided to support the localization of Pd from the contrast in the STEM image (Supplementary Figure 2.c). The net count signal of Pd and N on the write line also follows a similar geometric pattern. The alternative appearance of Pd along with a consistence N signal indicates the gap between island structure that would be the path for ions flux in the design of nanofluidic systems.

As shown in Supplementary Figure 3, we analyzed the TEM images to determine percentage of space occupied by the palladium islands for two devices with low and high island density, denoted as 'sparse islands' and 'dense islands' (the latter correspond to the device in Supplementary Figure 2). For dense islands, the coverage was found to be 70\%, whereas the coverage was only 45\% for the sparse islands. The image analysis pipeline consisted in local normalization\cite{sage2001easy}, followed by Gaussian blur and finally binarization at the mid-peak point of the bimodally distributed image. The images were acquired slightly away from the silicon
nitride aperture which ensures a continuous distribution of the palladium island.
\begin{figure}[!h]
	\centering
	\includegraphics[width=1\linewidth]{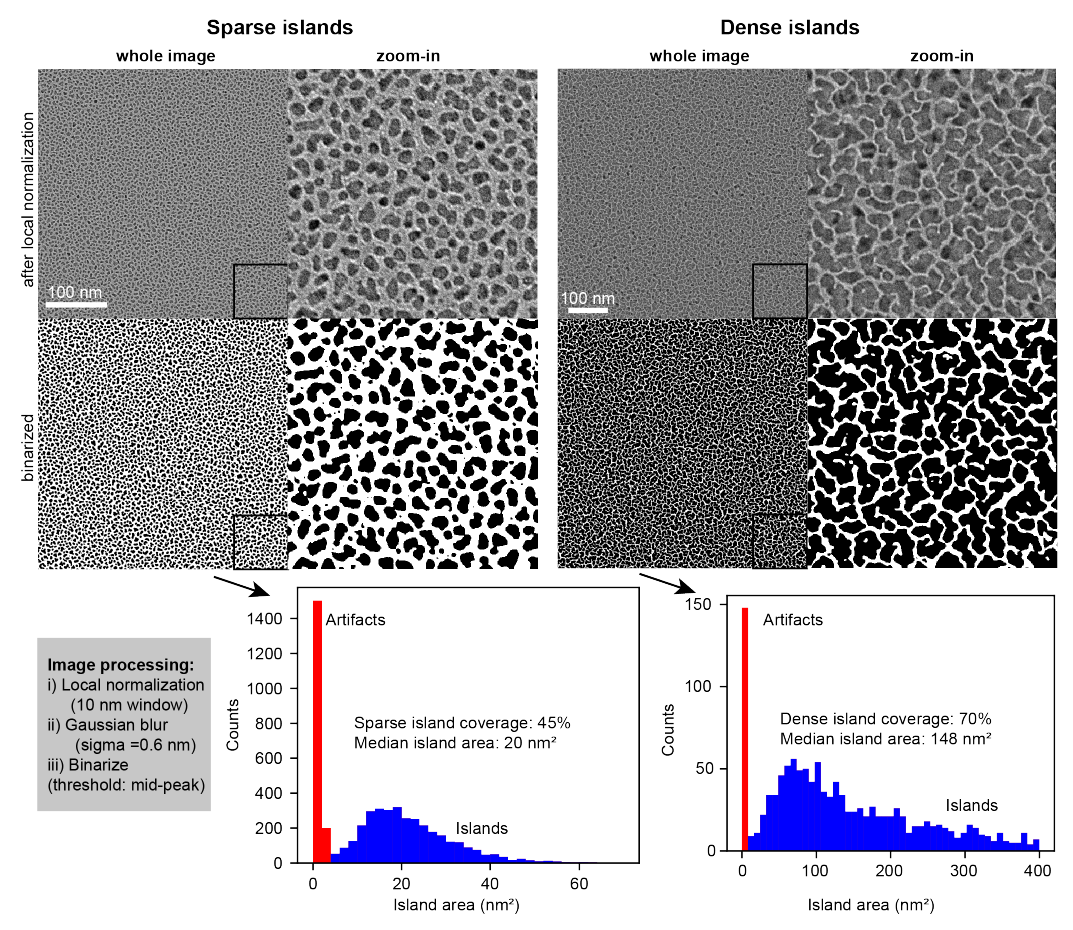}
	\caption{ \footnotesize	\textbf{Quantification of the palladium island coverage for two different islands density}. Starting from the locally normalized TEM images above, we obtain the binary images at the bottom left following the protocol described on the top right. \textit{ImageJ} was used for image processing, and local normalization was performed using the plugin Sage \& Unser\cite{sage2001easy}.  Following binarization, we obtain the island area distribution at the bottom right. The left peak (red, area < 4nm$^2$) corresponds to false positives in the detection of islands due to the imperfect binarization on noisy TEM images. We discard these small islands and count the island coverage by dividing the total area occupied by larger islands (blue, area > 4 nm2) by the image area, yielding a coverage of 45\%. However, for the dense island device, such as (Supplementary Figure 2), the coverage is up to 70\%.}
\end{figure}

\pagebreak

\subsection{Step 2:  2D material transfer}

The chosen materials are exfoliated on Si wafer with a nitride layer of 200 nm thickness.  A flat crystal with an area of minimum 20x20 microns is then selected. It is transferred using standard dry-transfer technique as detailed in methods section. The thin SiN window at the center of the chip with the hole in its middle constitutes the target. The only requirement here is that the hole in the center has to be covered.  The devices are then cleaned with acetone and isopropanol prior to the measurements.  supplementary Figure 4 shows images of finished devices with different top layer materials.

\begin{figure}[!h]
	\centering
	\includegraphics[width=1\linewidth]{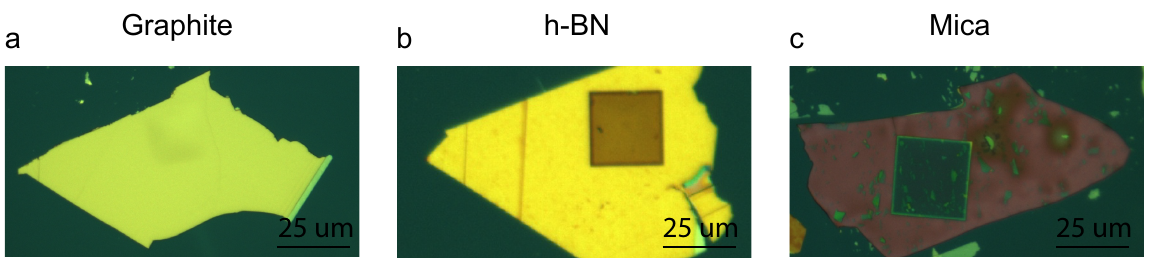}
	\caption{ \footnotesize	\textbf{Optical microscopy images of finished devices (a-c)}The Window can be observed in h-BN and Mica crystals which are transparent.}
\end{figure}

\pagebreak
\section{List of results per device}
Supplementary table 1 report all experimental results in this work. For each device, the corresponding available experiments are listed as well as correspond figures. We also provide On-state, Off-state, and conductance ration under sinusoidal periodic operation for the experimental conditions presented.

\scriptsize
\begin{center}
\begin{tabular}{ | m{1.5cm} | m{3cm}| m{1.5cm} | m{1.5cm}| m{1.5cm} | m{1.5cm}| m{1.5cm} | m{1cm}| } 
  \hline
 & Experiment (Figure or video) & Concentration & Species & Materials & $G_\text{on}$ (nS)&$G_\text{off}$ (nS)& $\frac{G_\text{on}}{G_\text{off}}$ \\ 
  \hline
  Device 1 &IV(F1), amplitudes(F2), charge(F2), pulses(SF7), frequencies (SF10)& 1M& KCl& Graphite&120&6&20 \\ 
  \hline
   Device 2  & Optics(F3 and video 1 )& 1M& KCl& Graphite&224&6&37  \\ 
  \hline
  Device 3  & Logic(F4)& 1M & KCl &Graphite&-&-&-  \\ 
  \hline
    Device 4  &Logic(F4) & 1M & KCl &Graphite &-&-&-  \\ 
\hline
    Device 5  & IV(SF5), pulses (SF6 and SF ) & 1M &KCl &Graphite &660& 31& 22\\ 
\hline
 Device 6  & IV(SF5), pulses (SF6 and SF17), stability (SF18) &  1M &KCl &Graphite&370&23&16  \\ 
\hline
 Device 7  & IV(SF5), frequencies(SF10), concentration(SF14) & 1mM-1M &KCl &Graphite&596&62&10  \\ 
 \hline
 Device 8  & IV(SF5), frequencies(SF10), species(SF13) &  1M& KCL, CaCl$_2$, AlCl$_3$ &Graphite&276&17&16   \\ 
  \hline
  Device 9  & IV(SF5) & 1M&KCl &Graphite&282&22&13  \\ 
  \hline
Device 10  & IV(SF5) & 1M&CaCl$_2$ &Graphite&350&96&4  \\ 
\hline
Device 11 & IV(SF5), amplitudes(SF8) &   1mM-1M (SF9)&CaCl$_2$ &Graphite&187&33&6  \\ 
\hline
Device 12 & IV(SF5), amplitudes(SF8), pulses (SF7 and SF12) &1M & CaCl$_2$ & Graphite&643&125&5 \\ 
\hline
Device 13 & amplitudes and waveform (SF9), stability (SF18)  &  1M&KCl &Graphite&606&64&9 \\ 
\hline
Device 14 & frequencies(SF10)  & 1M&KCl &Graphite&900&84&11 \\ 
\hline
Device 15 & pulses(SF12) &  1M&KCl &Graphite&-&-&- \\ 
\hline
Device 16 & concentration (SF14)  &  3mM-300mM&KCl &Graphite&422&15&26 \\ 
\hline
Device 17 & IV(SF15) &   1M&KCl &h-BN&154&23&7 \\ 
\hline
Device 18 & IV(SF15) &  1M&KCl &mica&300&125&2 \\ 
\hline
Device 19 & IV(SF15) & 1M&KCl &mica&155&115&1\\ 
\hline
Device 20 & pressure (SF16) &  1M&KCl &Graphite&-&-&- \\ 
\hline
Device 21 & pressure (SF16) &   1M&KCl &Graphite&-&-&- \\ 
\hline
Device 22 & concentration gradient (SF16) & 1M&KCl &Graphite&-&-&-\\ 
\hline
Device 23 & Optics (SF20 and video 2) &  1M&KCl &Graphite&550&17&33 \\ 
\hline
Device 24 & Optics (video 3)  & 1M&KCl &Graphite&595&10&60 \\ 
\hline
Device 25 & Optics (video 4) &  1M&KCl &Graphite&500&13&38 \\ 
\hline
Device 26 & Optics-LTP (video 5) &  1M&KCl &Graphite&-&-&-\\ 
\hline
Device 27 & Optics (video 6) & 1M&KCl &mica&220&200&1\\ 
\hline
Device 28 & Optics (video 7) &   1M&KCl &mica &160&80&2\\ 
\hline
Device 29& logic (SF26) &  1M&KCl &Graphite&-&-&- \\ 
\hline
Device 30& logic (SF27)  &  1M&KCl &Graphite &-&-&-\\ 
\hline
Device 31& logic (SF27)  & 1M&KCl &Graphite&-&-&- \\ 
\hline
Device 32& logic (SF28)  & 1M&KCl &Graphite&-&-&- \\ 
\hline

\end{tabular}
\end{center}
 \footnotesize{Supplementary table 1: }\textbf{List of devices with corresponding experiments.} F is for Figure and SF is for Supplementary Figure. 
\normalsize

\normalsize

\section{Nanofluidic measurements with a single device}
\subsection{IV sweeps}
To demonstrate HACs' reproducibility and robustness we show IV sweeps on 8 different devices (Supplementary Figure 5) in addition the the one presented in the main text. It can already be noticed that the memory effect is stronger with KCl than CaCl$_2$. This is quantified in section 3.9.

\begin{figure}[!h]
	\centering
	\includegraphics[width=1\linewidth]{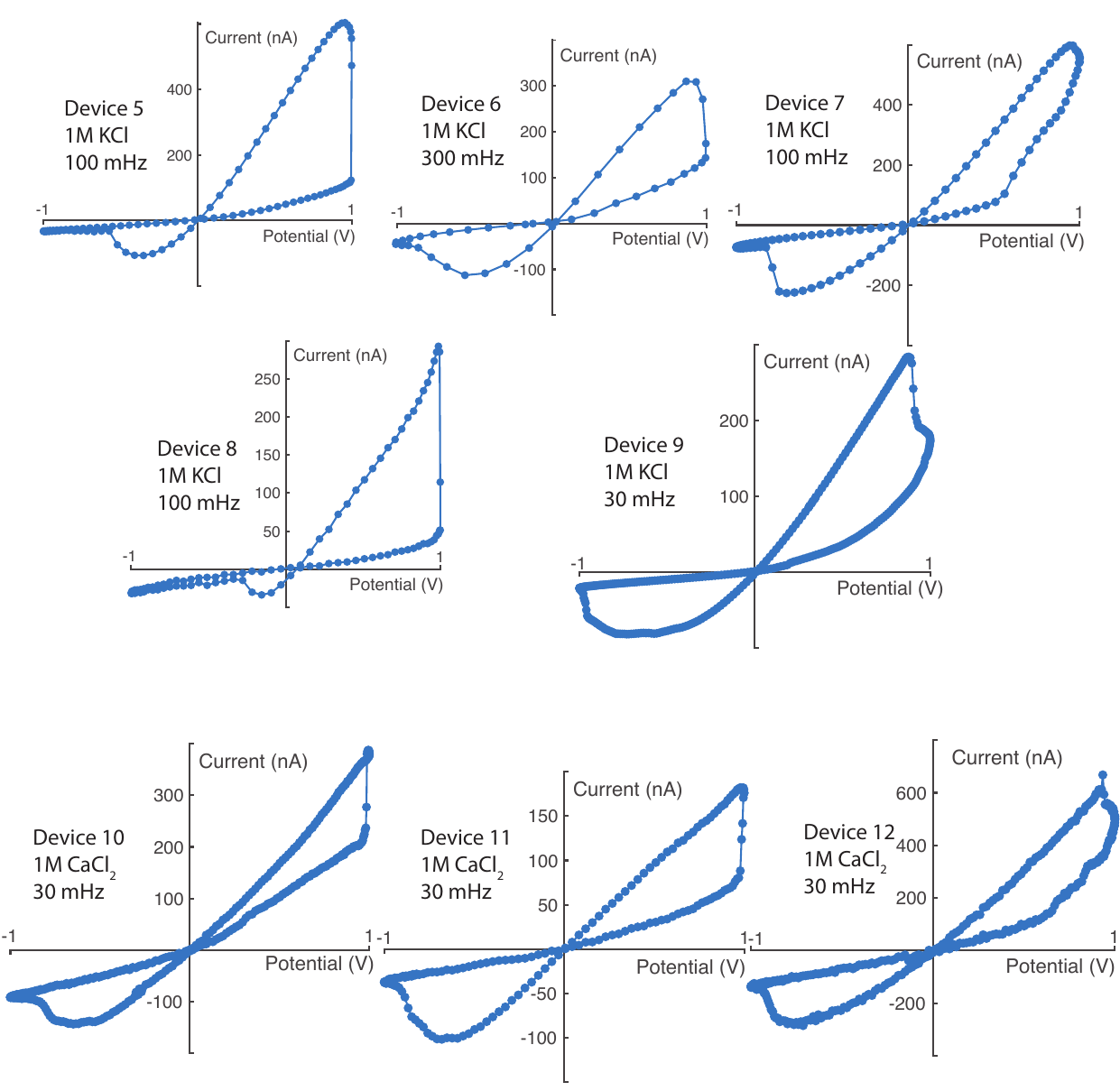}
	\caption{ \footnotesize	\textbf{IV sweeps.} Each graph is obtained with a different device. The frequency and salt is indicated for each device. For all devices the salt concentration is 1M and the top layer material is graphite.}
\end{figure}
\clearpage

\subsection{Pulse programming}

To check HACs' ability for neuromorphic computing we perform basic pulse programming operation (Supplementary Figure 6). The device is reset with a high amplitude negative programing pulse and set with a high amplitude positive pulse.  The conductance can be measured with a low amplitude positive read pulse. The device can be reset or set with 2s programming pulse and read pulse in ON state is around five times larger than read pulse in OFF states.

\begin{figure}[!h]
	\centering
	\includegraphics[width=1\linewidth]{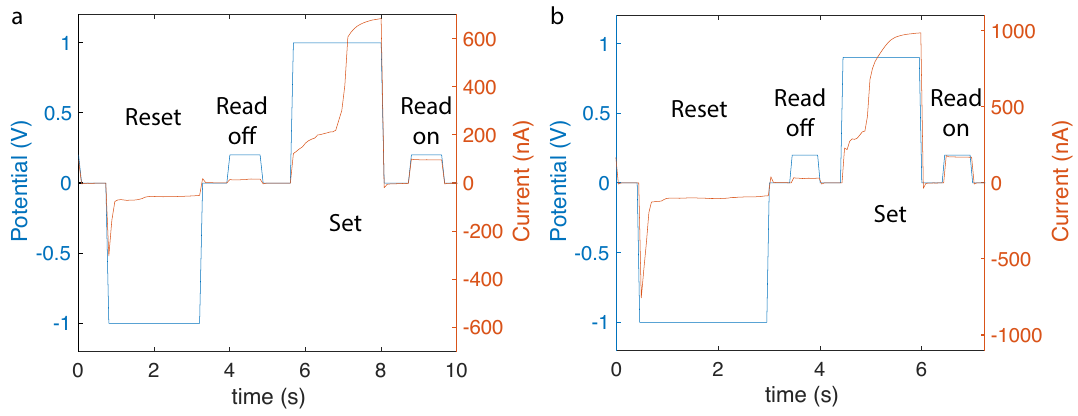}
	\caption{ \footnotesize	\textbf{Programing pulses. a,} Device 6. \textbf{b,} Device 5. The salt used is KCl at a concentration of 1M.}
\end{figure}

\subsection{Setting at intermediate levels of conductance}
We check that HACs can be set to intermediate level of conductance by applying successive short voltage pulses (0.5-1s). Device are then reset progressively with negative voltage pulses of same duration. We observe that it is indeed possible to achieve intermediate level of conductances as shown in Supplementary Figure 7 thereby mimicking progressive synaptic potentiation and depression. This allow in principle the implementation of in-memory computing algorithm emulating more complex synaptic behavior like spike-timing dependent plasticity (STDP).

\begin{figure}[!h]
	\centering
	\includegraphics[width=1\linewidth]{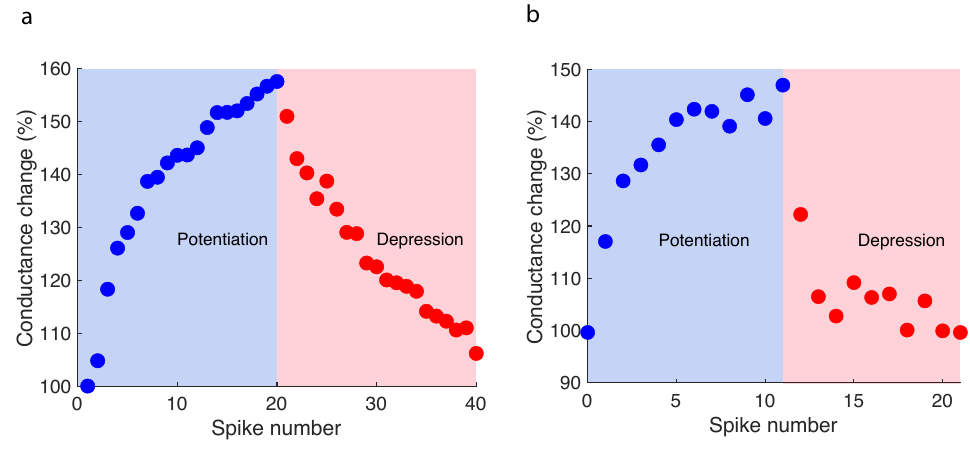}
	\caption{ \footnotesize	\textbf{progressive potentiation and depression of HACs through intermediate conductance levels. a},  (Device 1-1M KCl). The conductance is increased (decreased) with programming pulses of +1V (-1V) with duration of 0.5s. Between each pulse it is accessed with non intrusive read pulses of 1s and 100mV. 
(Device 12-1M CaCl2). The conductance is increased (decreased) with programming pulses of +1V (-1V) with duration of 1s. Between each pulse it is accessed with non intrusive read pulses of 5s and 100mV. \textbf{b}, (Device 12-1M CaCl2). The conductance is increased (decreased) with programming pulses of +1V (-1V) with duration of 1s. Between each pulse it is accessed with non-intrusive read pulses of 5s and 100mV.
}
\end{figure}
\pagebreak

\subsection{Additional data for sweeps at with different amplitudes}
We provide in Supplementary Figure 8 additional data with applied sinusoidal bias of varying amplitude. We can notice, as for device A in Figure 2, that the voltage at which the switching occurs is dependent on the bias amplitude. 

\begin{figure}[!h]
	\centering
	\includegraphics[width=1\linewidth]{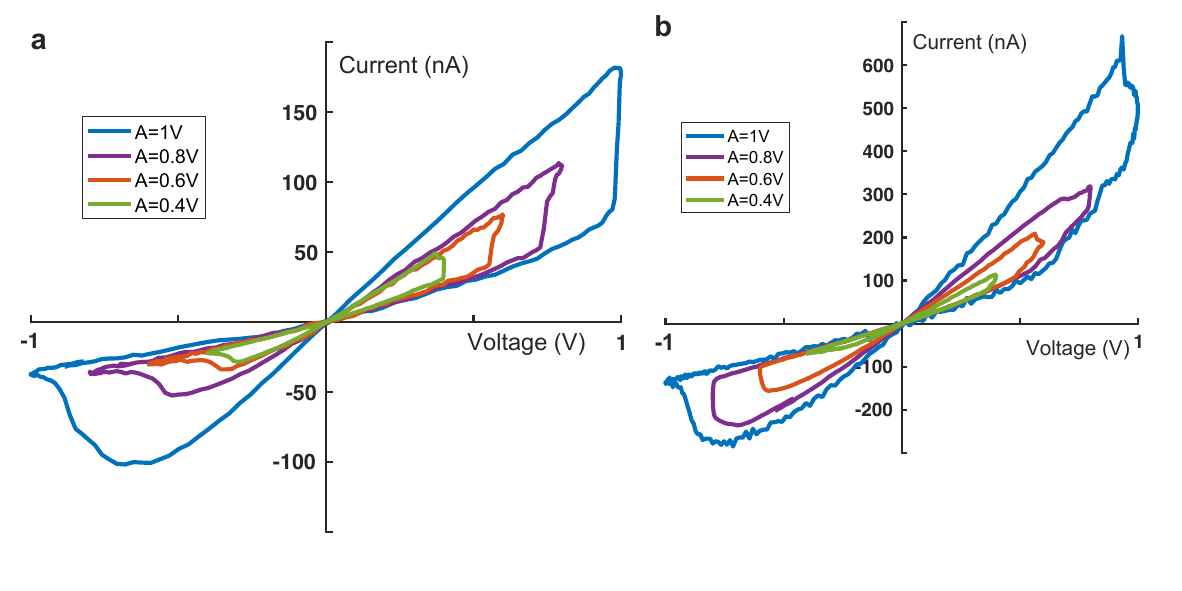}
	\caption{ \footnotesize	\textbf{Sweeps with different amplitudes for two supplementary devices} \textbf{a,} Device 11, \textbf{b,} Device 12. The salt used is CaCl$_2$ at 1M and the frequency is 30 mHz. }
\end{figure}
\pagebreak

\subsection{Data for sweeps with different voltage waveform}

To verify if the effect had a strong dependance on the applied waveform shape, we compare in Supplementary Figure 9 sweeps at different amplitude using triangular and sinusoidal wave. While sinusoidal wave result in a slightly larger hysteresis as more time is spent at large voltage, we find that the waveform shape has little effect on the threshold.

\begin{figure}[!h]
	\centering
	\includegraphics[width=1\linewidth]{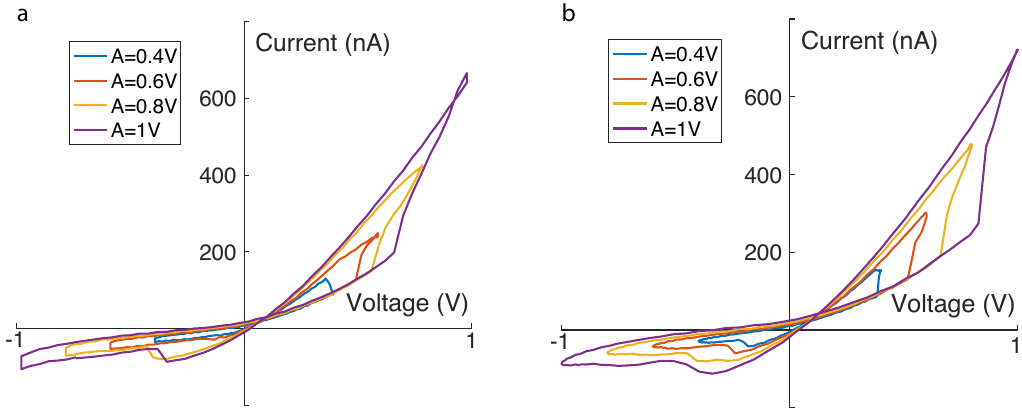}
	\caption{ \footnotesize	\textbf{Current-voltage (IV) curves at different amplitudes using triangular waves. (\textbf{a}) and sinusoidal waves (\textbf{b}).} Device 13 at 1M KCl. 100 mHz. 
}
\end{figure}

\subsection{Additional data for sweeps at different frequencies}

We provide in Supplementary Figure 10 additional data with applied sinusoidal bias of varying frequencies. At highest frequencies, the memristive effect disappears, leading to a ohmic behavior. Small ellipsoid can be observed because of capacitive effects. At lowest frequencies, the IV characteristic is diode-like (Supplementary Figure.10-d). We notice that slowing at low frequencies (below 10 mHz) seems to damage the device as recovering the OFF-state is not possible. Thus, application of high voltage for prolonged time should be avoided. This is however not required as memristors are expected to operate as fast as possible.

\begin{figure}[!h]
	\centering
	\includegraphics[width=1\linewidth]{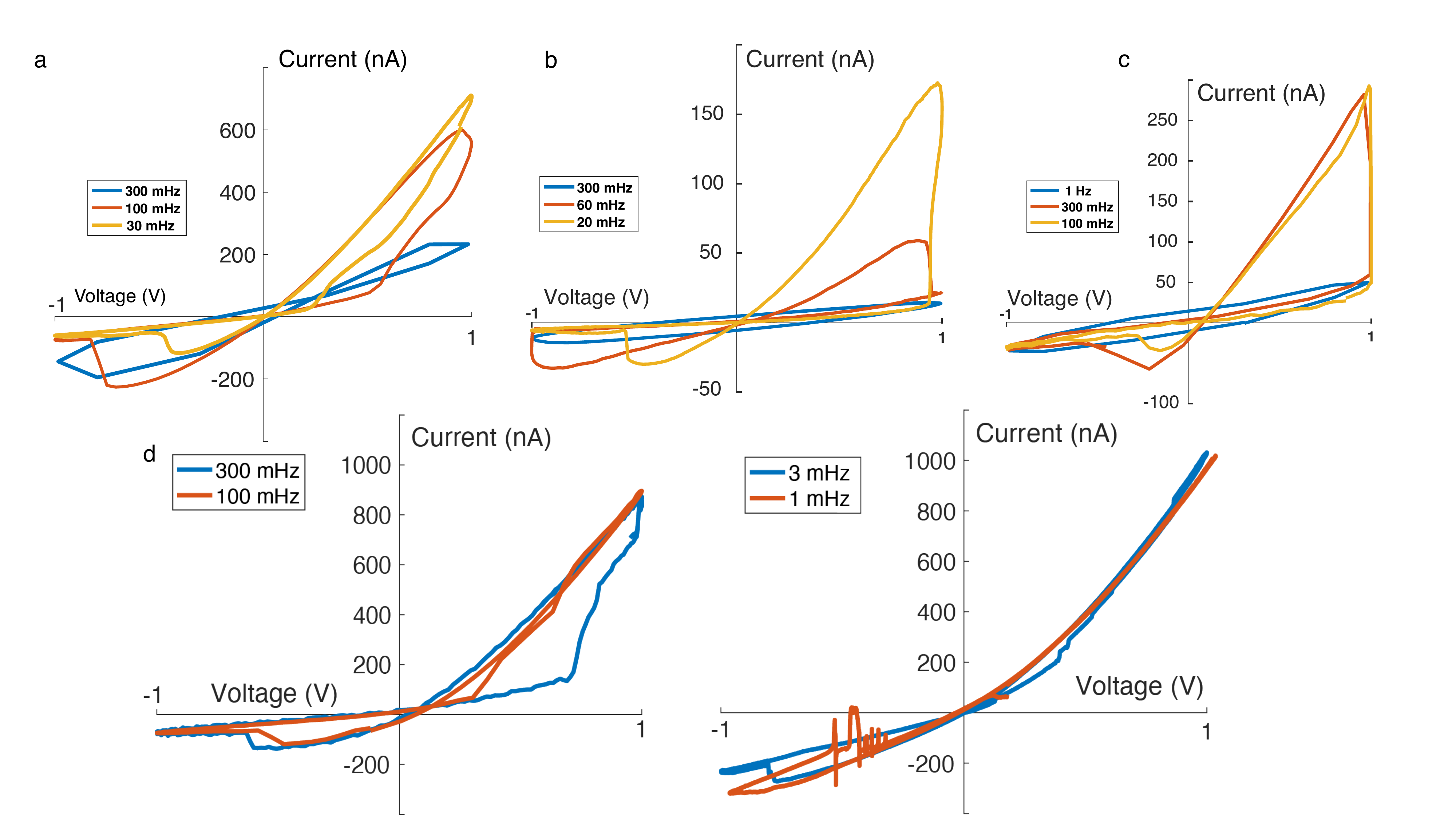}
	\caption{ \footnotesize	\textbf{Sweeps with different amplitudes for two supplementary devices.} \textbf{a,} Device 7, \textbf{b,} Device 1, \textbf{c,} Device 8. \textbf{d}, Device 14. The salt used is KCl at 1M. }
\end{figure}

\pagebreak

\subsection{Additional data for the charge threshold}
Following the same procedure as in main text Figure 2.b-c, we reported the charge threshold for three additional devices (Supplementary Figure 11.a-d). We find that the charge threshold varies by maximum three for a given device and is comprised in the 30-300 nC range for the four studied devices. 

\begin{figure}[!h]
	\centering
	\includegraphics[width=1\linewidth]{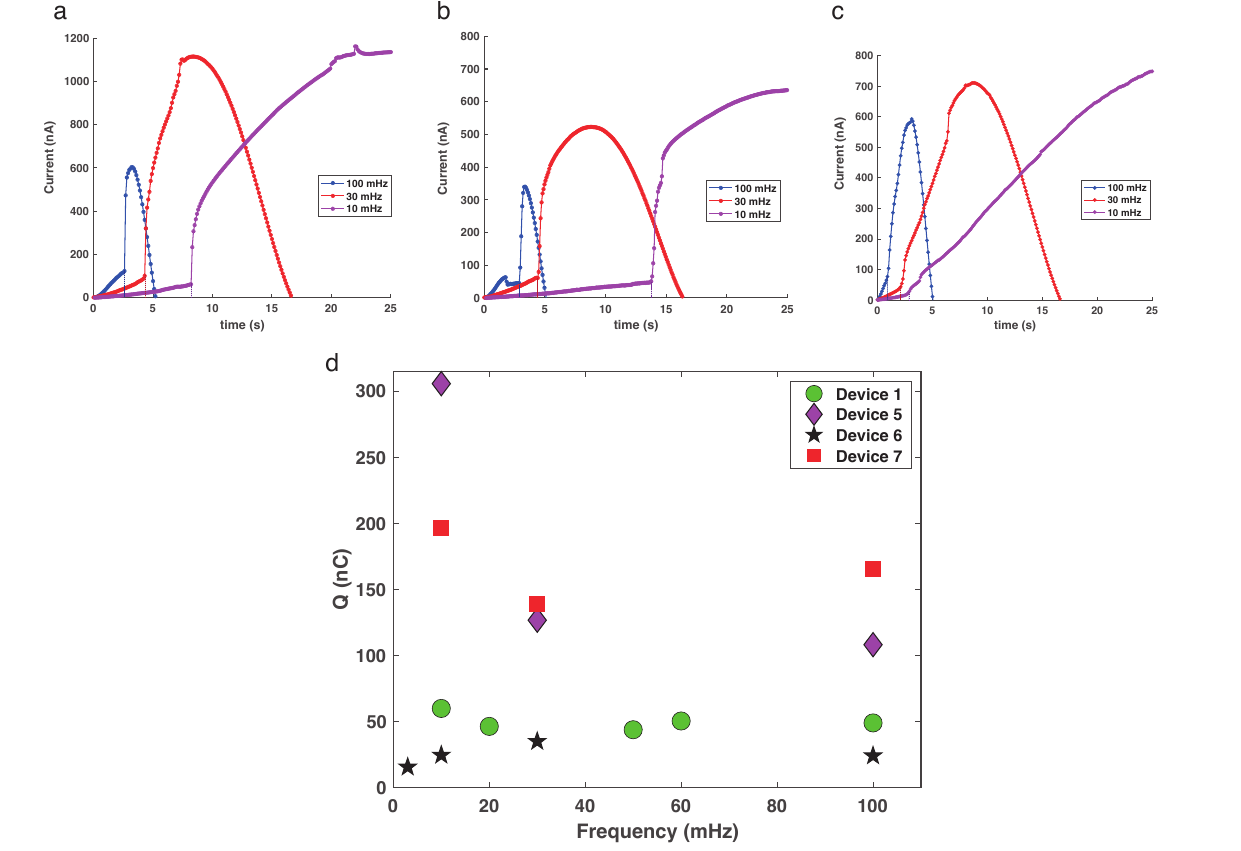}
	\caption{ \footnotesize	\textbf{Additional data for charge threshold (1M KCl). a-c,} Setting of the device with sinusoidal potential of various frequencies for devices 5-7 respectively. The vertical dashed lines indicate the threshold time for each frequency. \textbf{d,} Charge threshold, Q, for various frequencies and devices. Device 1 is presented in the main text.}
\end{figure}
\pagebreak

\subsection{Memory retention}

We check how the conductance evolve after setting/resetting in Supplementary figure 12. To do so we set/reset the device with positive/negative voltage pulses and access the conductance state with a weakly disturbing voltage pulse train of low amplitude oscillating around 0. We find that even if the conductance can experience an overshoot after setting/resetting (short-term potentiation) it does stabilize at a level higher (lower in case of reset) than its pre-pulse level. HACs thus indeed exhibit long-term potentiation. We repeat this experiment with optical feedback in section 4.

\begin{figure}[!h]
	\centering
	\includegraphics[width=1\linewidth]{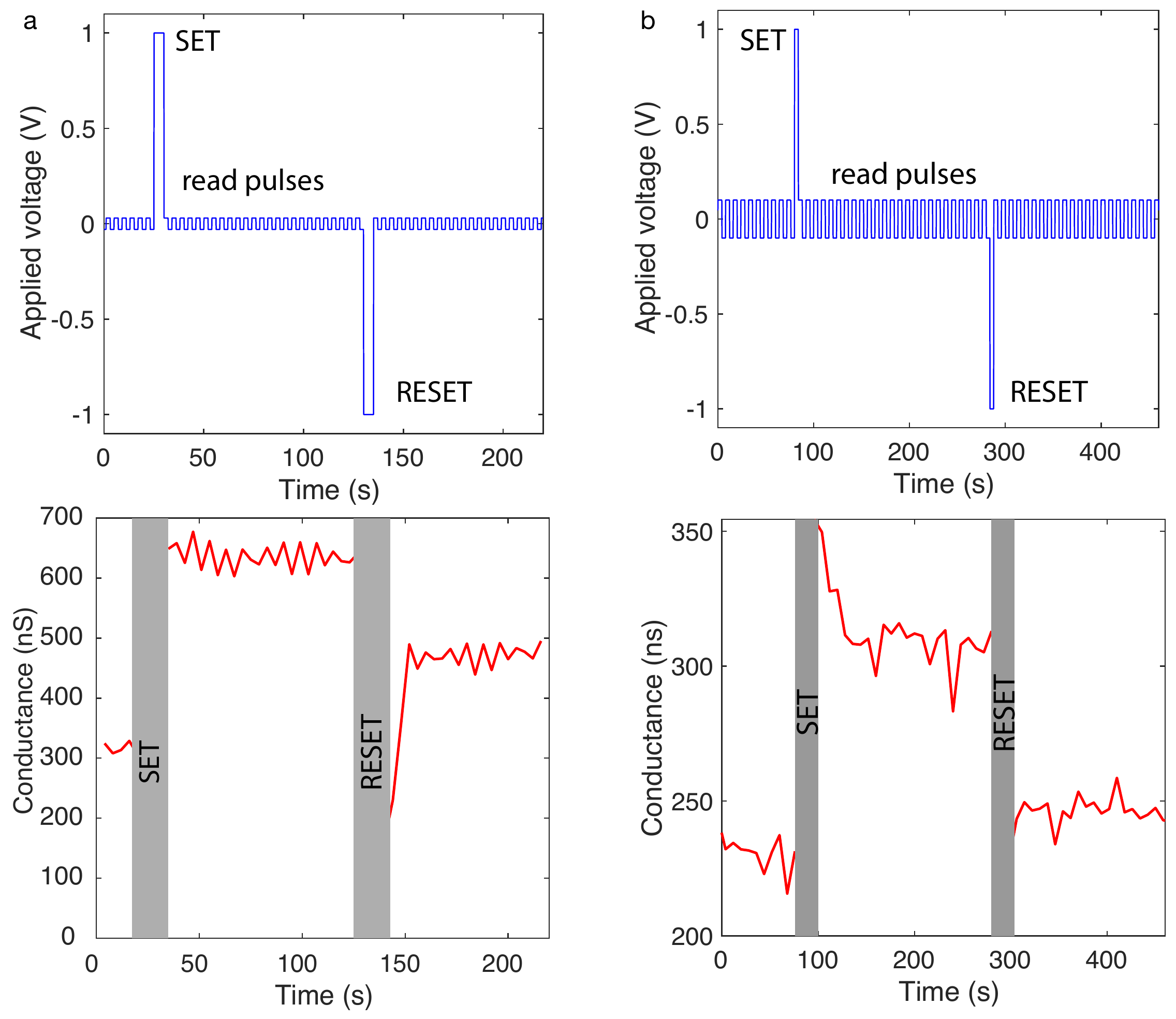}
	\caption{ \footnotesize	\textbf{Implementation of long term potentiation with two devices. a}, Electrokinetic data (Device 15-1M KCl). The applied voltage is displayed on the top and the extracted conductance is shown on the bottom. The conductance is measured with voltage pulse train oscillating between $\pm$ 30 mV. The device is set/reset with a single positive voltage pulse of +/-1V and a duration of 5 seconds. \textbf{b}, Electrokinetic data (Device 12-1M CaCl$_2$). The applied voltage is displayed on the top and the extracted conductance is shown on the bottom. The conductance is measured with voltage pulse train oscillating between $\pm$ 100 mV. The device is set/reset with a single voltage pulse of $\pm$1V and a duration of 5 seconds.}
\end{figure}
\pagebreak

\subsection{Dependency with ionic species}

The memory effects increases with decreasing ionic valence and is maximum for KCl (Supplementary Figure 13.a-b).

\begin{figure}[!h]
	\centering
	\includegraphics[width=1\linewidth]{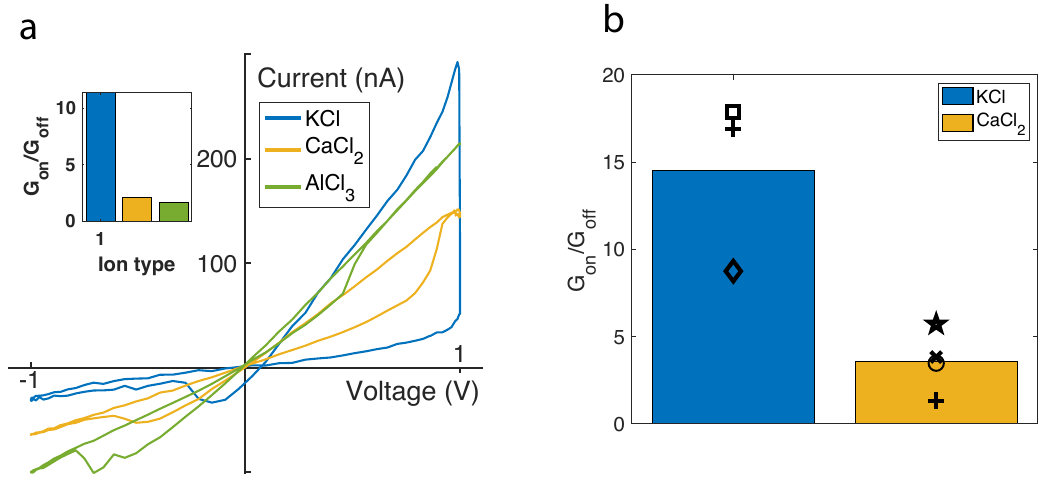}
	\caption{ \footnotesize	\textbf{Dependency of the effect with ion type (at a concentration of 1M). a,} IV characteristics with different salts for device 8. The frequency is 100 mHz. Inset: Extracted conductance ratio for the three salts. \textbf{b,} Comparison of conductance ratios between monovalent and divalent salts with 6 different devices. Each symbol represents a device. The frequency is 30 mHz.  }
\end{figure}

\subsection{Dependency with the salt concentration}

The memristor effect can be observed in HACs at concentrations higher than 100mM (Supplementary Figure 14.a-b). 
It can be noted that HACs are weakly conductive at concentration below 30-100mM (Supplementary Figure 14.a-c). Therefore the conductive, high-concentration regime, is where the memory can be observed.

\begin{figure}[!h]
	\centering
	\includegraphics[width=1\linewidth]{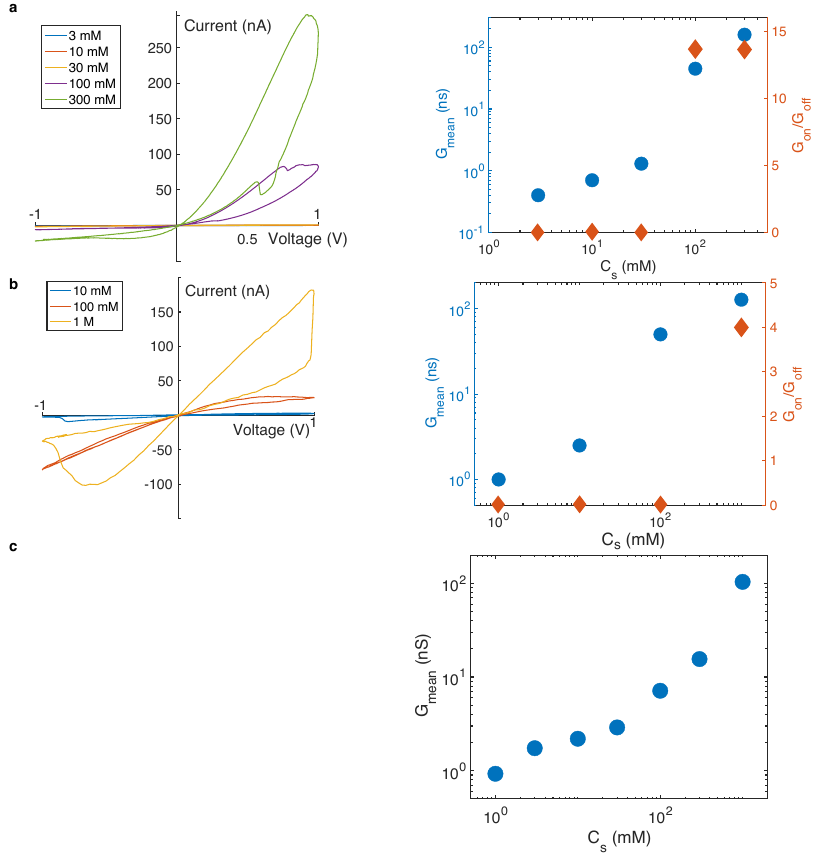}
	\caption{ \footnotesize	\textbf{HAC's conductance and conductance ratio with the salt concentration. a,} Device 16-KCl. Left: IV characteristics for different salt concentrations at 10mHz. Right: quasi-static mean conductance (Blue points) and conductance ratio measured at 10 mHz (red diamonds) at different salt concentration.  If the memristive effect cannot be observed, the conductance ratio is defined as 0. \textbf{b,} Device 11-CaCl$_2$. Same panels and legend as \textbf{a}. \textbf{c,} Quasi-static mean conductance vs concentration for device 7-KCl.   }
\end{figure}

\clearpage

\subsection{Dependency with type of 2D material}

We changed the top layer material to further characterize the memory effect in HACs. We find that while the effect can still clearly be observed with a top layer made of h-BN, the IV characteristics with a mica top is quasi-linear (Supplementary Figure 15). This indicates that the effect is controlled by the surface charge coming from the top layer as mica bears a weaker surface charge relative to h-BN or graphite.

\begin{figure}[!h]
	\centering
	\includegraphics[width=1\linewidth]{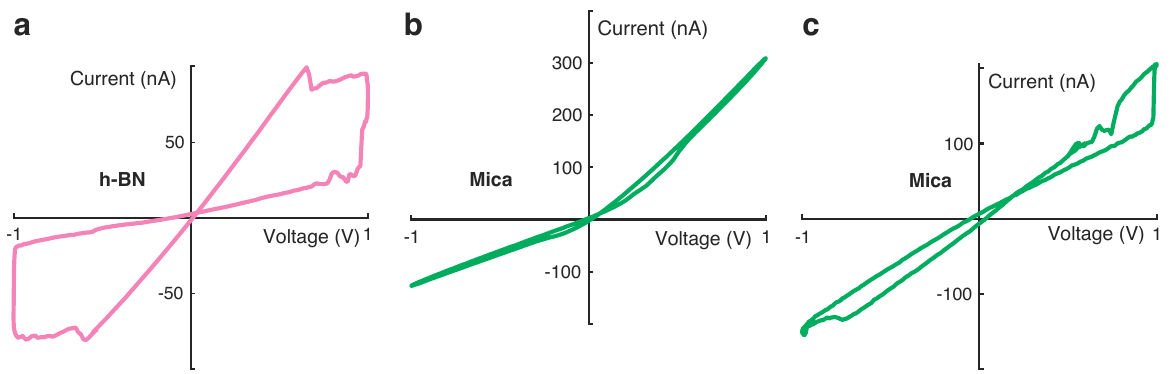}
	\caption{ \footnotesize	\textbf{Dependency of the top material. (1M KCl-30 mHz) a,} IV sweep for a device with h-BN cap (Device 17) \textbf{b-c,} Device sweeps for two devices with mica cap (Devices 18 and 19). }
\end{figure}

\subsection{Selectivity: streaming measurements}

To verify that our device are indeed selective, we performed pressure-driven streaming currents measurements as well as osmotic currents measurements under salinity gradients (Supplementary figure 16). Under the application of pressure, ions are moved by advection generated by fluid-flow. While bulk-ions do not contribute to the current as cations and anions compensate for each other, excess counter-ions generate an ionic "streaming" current. Thus measuring a streaming current shows channel selectivity and its sign is related to the sign of the net surface charge. In our case, we measure positive currents while applying pressure at the working electrode reservoir which is the one where the 2D crystal and large entrance are. This is consistent with the measured osmotic currents under salinity gradients. Altogether, these results point to positive counter-ions (K$^+$) screening a negative net surface charge.

\begin{figure}[!h]
	\centering
	\includegraphics[width=1\linewidth]{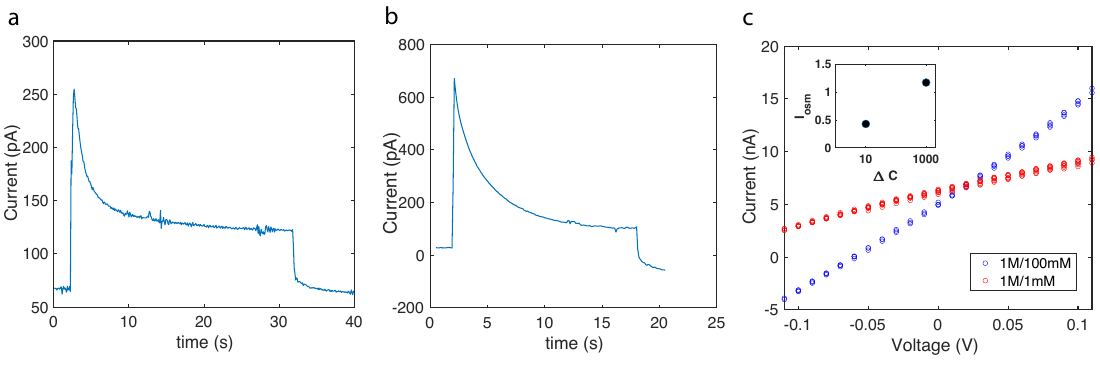}
	\caption{ \footnotesize	\textbf{Selectivity control with pressure-driven ionic transport. a,} Device 20, 1M KCl. The applied pressure is 500 mbar. \textbf{b,}  Device 21, 1M KCl. The applied pressure is 500 mbar. \textbf{c}, IV curve of HAC at two different concentration gradients (Device 22).  Inset: extracted osmotic currents. The osmotic currents are calculated as in Emmerich et al.\cite{emmerich2022enhanced} by subtracting the Nernst current to the current at V=0. }
\end{figure}

\pagebreak
\subsection{Setting with pulses of different magnitude and voltage pulses endurance tests}

The IMP logic requires a switching threshold to be achieved correctly. In term of pulse programming, a setting threshold means that if a pulse is not high or long enough to reach the charge threshold, the conductance is weakly affected ($V_{Q}^\beta$ in the IMP logic). On the other hand, If the pulse is sufficient to reach the charge threshold, the conductance will be strongly affected ($V_{Q}^\alpha$ in the IMP logic). Thus, even a small different of setting voltage can result in a large difference in output conductance. To verify that HACs indeed behave this way, we apply different setting voltages and monitor the change in conductance (supplementary Figure 17.a). We find that HACs are setted only for a voltage higher than a given value, as expected given the threshold dynamic and as required for IMP logic (supplementary Figure 17.a-b). We repeat this process four times and notice little variation (supplementary Figure 17.c). This demonstrates the stability of our devices which is also required to perform IMP logic where 8 setting/operation pulses are applied on each HAC.  

\begin{figure}[!h]
	\centering
	\includegraphics[width=1\linewidth]{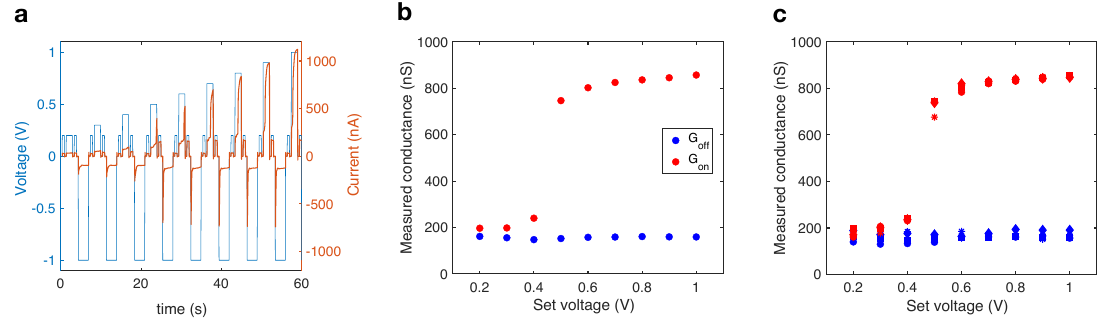}
	\caption{ \footnotesize	\textbf{Dependance on the magnitude of the programming pulse and stability-Device 5, 1M KCl. a,} Applied potential and resulting current (raw data). The conductance is measured with read pulses of 0.2 V and 0.5 s before and after the programming pulses. The programming pulses have a duration of 1.5s. Current during programming pulses exhibit an abrupt increase when the charge threshold is reached for programming pulses of voltages of 0.5V and higher like in the $\alpha$-case of Figure 4.d. \textbf{b,} Conductance measured with read pulses before and after each programming pulses, extracted from \textbf{a}.  This device is set for programming pulses larger thane 0.4V.  \textbf{c}, Repetition of \textbf{a-b} four times. Each symbol represents an iteration.}
\end{figure}

\pagebreak
\subsection{Endurance tests under sinusoidal potential}

We checked the endurance of HACs  using multiple applied sinusoidal voltage cycles  We observe excellent stability over 7 cycles (Supplementary Figure 18.a). For a larger number of cycles, the IV curves can drift. Yet we could observe up to 300 cycles (supplementary Figure 18.b).

\begin{figure}[!h]
	\centering
	\includegraphics[width=1\linewidth]{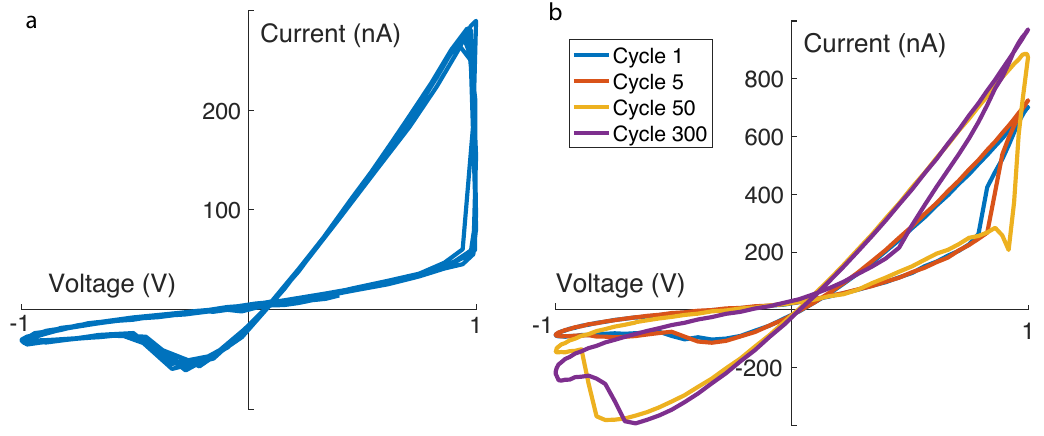}
	\caption{ \footnotesize	\textbf{Endurance test with sinusoidal applied bias. a,} IV-characteristics for 7 consecutive cycles. Device 5, 1M KCl-300 mHz. \textbf{b}, A few extracted individual cycles among 300 repetitions. Device 13, 1M KCl-100 mHz. }
\end{figure}

\pagebreak

\section{In operando optical measurements}
\subsection{Setup}

We performed in-operando observations using the setup sketched in supplementary Figure 19, where a water-dipping objective is used to image the device in a custom fluidic cell. As nanofluidic chips are smaller than the water-dipping objective, we designed a PEEK adapter (square with a 1 mm aperture) on which the chip was glued using epoxy resin (Araldite standard). We placed the chip upside down (graphite was facing down) to image inside the channels through the SiN window of the chip. This adapter was sandwiched between the two parts of the PEEK fluidic cell, held together by threads not shown on the sketch and sealed with O-rings. The bottom cell chamber had a leak-proof opening for the bottom electrode. The top cell chamber is open and has both the objective and the top electrode immersed. The objective used (Nikon CFI Plan 100XC W) has a working distance of 2.5 mm. The objective-sample distance was adjusted first coarsely using a z-positioner (conoptics M102A), and the sample was scanned using first a manual xy translation stage (Newport M-401). The optimal focus was found using finer motion control with a 3-axis piezoelectric stage (Thorlabs NanoMax 300).
\subsubsection{Microscope details}
The illumination light source was a 635nm continuous wave laser (Thorlabs HLS635). Its $som$1mW output was attenuated with ND filters (Thorlabs, E03/E04) to fill reach adequate illumination power densities. The gaussian beam size was enlarged in two steps: first through coupling the fiber output with a parabolic mirror (Thorlabs RC08FC-P01, 8 mm beam size output) and then by passing through a 2X adjustable beam expander (Thorlabs GBE02-A). Light was focused onto the back focal plane of the objective with an achromatic doublet lens (Thorlabs AC254-125-A-ML, 125 mm focal length). The objective was mounted at the bottom of a filter cube mount (Thorlabs DFM1L) which contained a 50/50 non-polarizing beam splitter (Thorlabs BS033), which directed 50\% of the illumination light towards the sample and let through 50\% of the light reflected at the sample towards the camera path. The image was formed on the camera (Hamamatsu ORCA Flash 4.0) using an achromatic doublet lens with a 200 mm focal length (Thorlabs AC254-200-A-ML).
\subsubsection{Image acquisition and post-processing}
The camera was controlled using Micro-Manager. The exposure time was set to 100ms to match the sampling rate of the electrokinetic measurements. The light intensity was adjusted to fill the camera dynamic range. The pixel size of 6.5 $\mu$m yielded a size of 65nm on the object plane, and a 4x4 binning was used in the acquisition to compress the images ($\sim$260 nm pixel size on the final image). Digital image rotation was applied when needed, and to remove laser speckle effects, images were denoised with a second-order low-pass Butterworth filter of cutoff frequency 0.05. Image processing was done using the Python library scikit-image.

\begin{figure}[!h]
	\centering
	\includegraphics[width=1\linewidth]{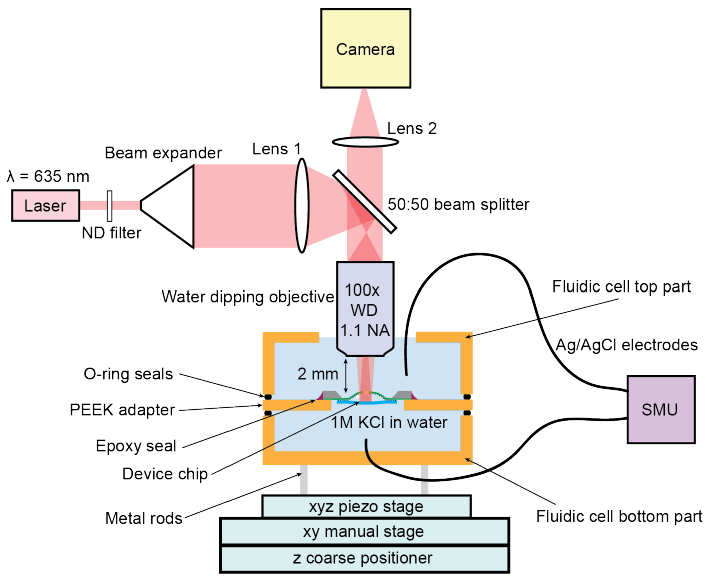}
	\caption{ \footnotesize	\textbf{Setup for \textit{in operando} optical measurements}}
\end{figure}

\pagebreak

\subsection{Additional Devices}

In addition to the device presented in Figure 3 (Supplementary Movie 1) of the main text, we have measured six supplementary devices. Four are made with a graphite top layer (Supplementary Movie 2-5) and two are controls using a mica top layer (Supplementary Movie 6-7). One graphite device shows threshold mechanism similar as the one presented in main (Device 23). For the two other graphite devices (Devices 24 and 25), the blister forms directly at the pore location and the threshold is therefore attributed at some abrupt dynamic taking place outside of the SiN window and therefore the field of view. 

\begin{figure}[!h]
	\centering
	\includegraphics[width=1\linewidth]{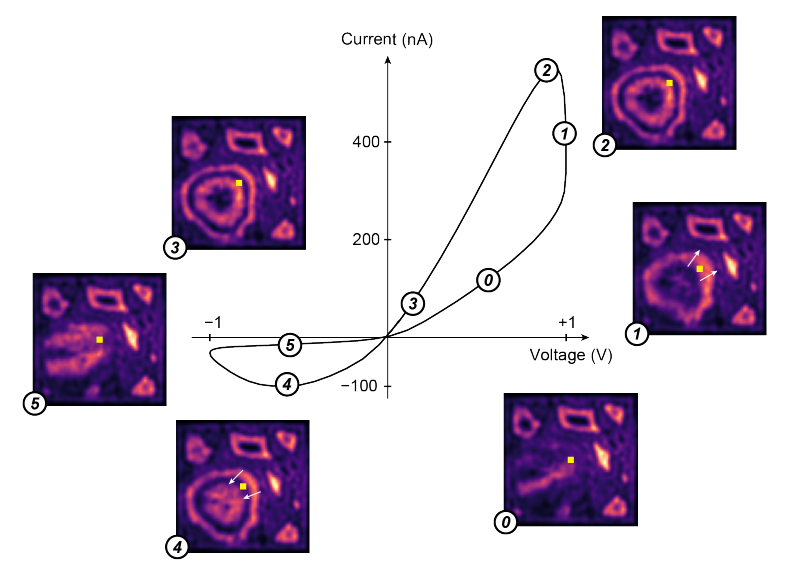}
	\caption{ \footnotesize	\textbf{IV characteristic at (1V, 300 mHz, device 23) with optical images of the SiNx window at different time points marked by white dots and numbers.} White arrows illustrate blister motion and yellow squares mark the pore region.}
\end{figure}

\subsection{Memory retention \textit{in operando} }

We performed \textit{in-operando} memory retention experiment as was done in section 3. We find that the diminution of the conductance and its stabilization after setting is consistent with the evolution of the blister (Supplementary Figure 21 and Supplementary Movie 5). This demonstrates that the ability to mimic long-term potentiation exhibited by HACs is related to their blister dynamics.

\begin{figure}[!h]
	\centering
	\includegraphics[width=1\linewidth]{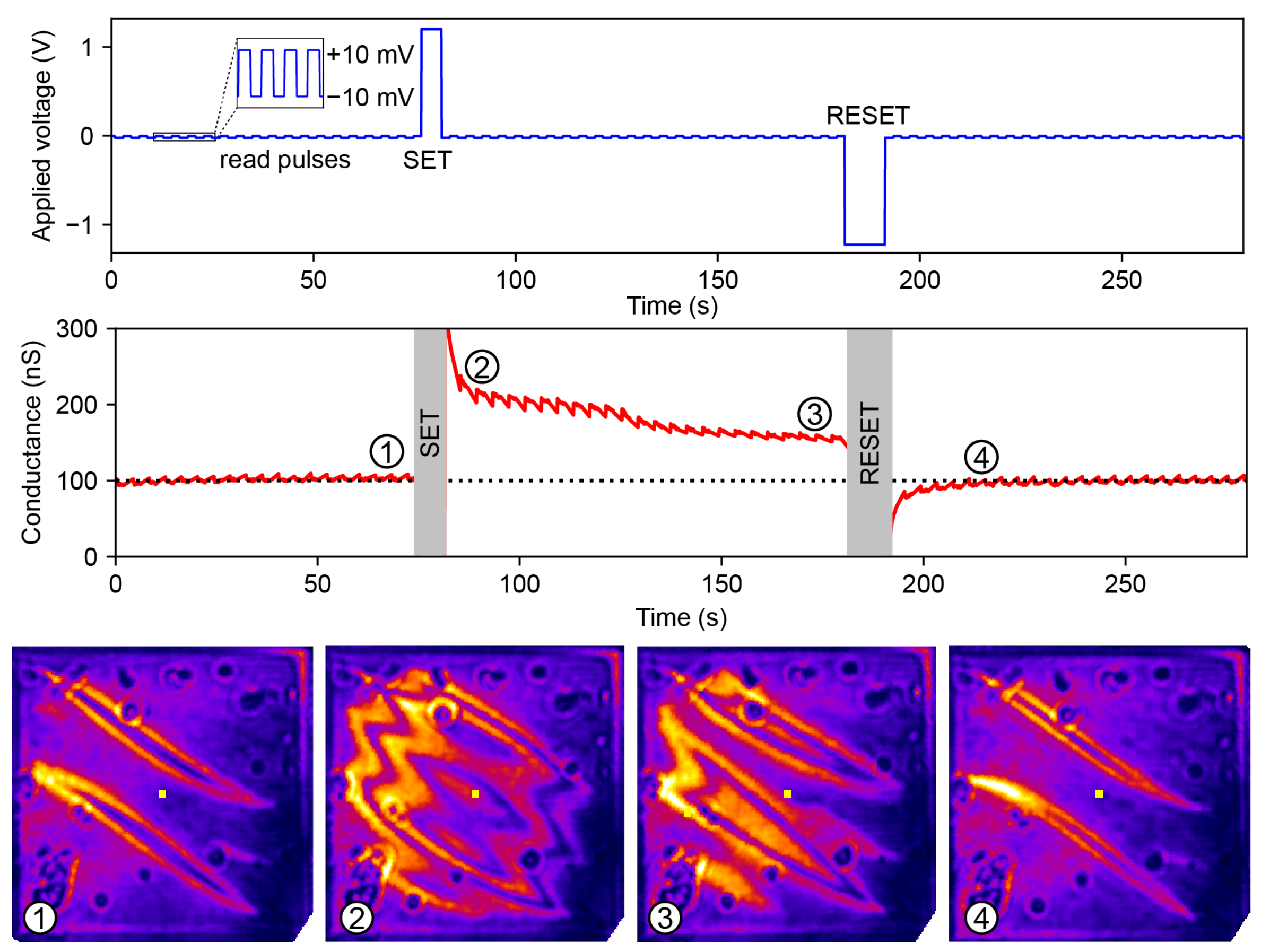}
	\caption{ \footnotesize	\textbf{Memory retention experiment with optical feedback (Device 26-1MKCl)}. Applied voltage (top), resulting conductance (middle) and snapshots taken at times indicated in conductance panel (bottom). After setting, the blister recedes before stabilizing in agreement with the ON-state conductance. The reset pulse results in the blister shrinking and thus the return of the conductivity to the OFF-state value. The yellow square indicates the pore region.}

\end{figure}

\subsection{Mica controls \textit{in operando} }

To confirm our hypothesis regarding the combined role of surface charge and adhesion of graphite to drive the memristive mechanism, we performed \textit{in-operando} optical observations with two mica controls.  As can be noticed in supplementary figure 22, there are some blisters in the vicinity of the SiN aperture. The main difference with graphite here is that the application of voltage has a very limited impact on the dynamics of these blisters. Interestingly, the limited blister formation observed for mica devices consistently appeared at negative voltages (outward pointing electric field), which seems to imply that these devices respond to anion motion rather than cation. We attribute this difference to a lower $\Sigma^2/\Gamma$ of mica resulting in marginal mechanical deformations (Supplementary videos 6-7). Thus, mica-caped devices remain in an intermediate state of conductance with limited variation of their conductance. This intermediate level of conductance was observed consistently on four devices.

\begin{figure}[!h]
	\centering
	\includegraphics[width=1\linewidth]{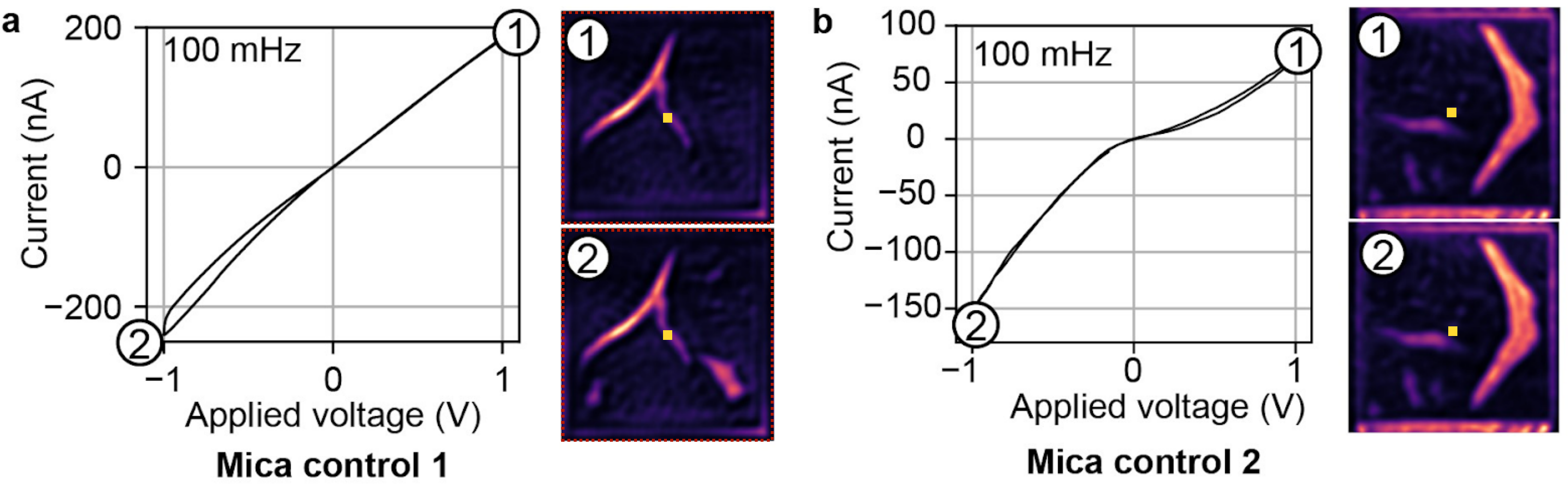}
	\caption{ \footnotesize	\textbf{Optical and electrokinetic results for two mica-caped devices under applied sinusoidal bias}. 1M-KCl-100 mHz. Yellow squares represent pore region. Device 27 (\textbf{a}) and  28 (\textbf{b}).}

\end{figure}

\subsection{Endurance tests under sinusoidal potential \textit{in-operando} }

We observed optically the endurance of HACs \textit{in-operando} using multiple applied sinusoidal voltage cycles. We find excellent stability of the memristive hysteresis as well as blister dynamics over 17 cycles (supplementary Figure 23). 
\begin{figure}[!h]
	\centering
	\includegraphics[width=1\linewidth]{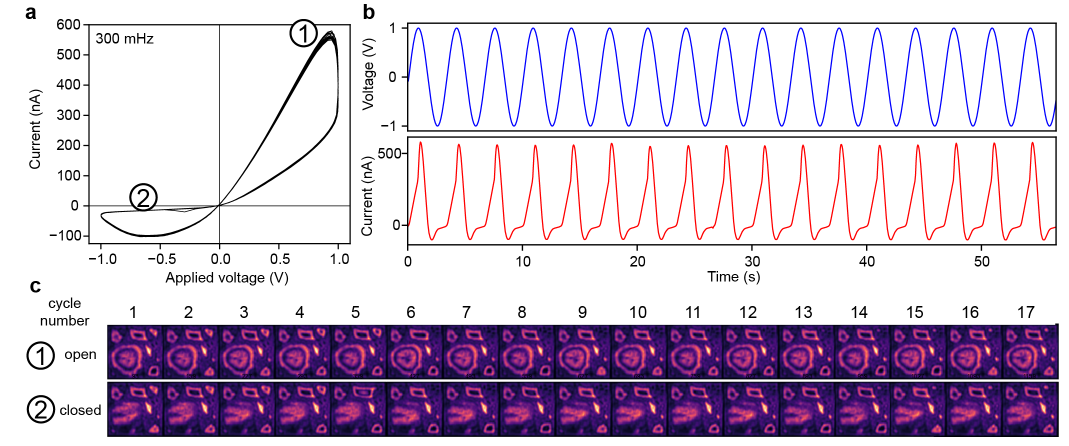}
	\caption{ \footnotesize	\textbf{In-operando measurements of 17 cycles-Device 23, 1M KCl. a,} IV characteristics of the 17 cycles. \textbf{b}, Applied voltage and resulting current in time for the 17 cycles. \textbf{c}, Snapshot extracted for each cycles corresponding to points in \textbf{a}.}
\end{figure}

\pagebreak

\section{Theoretical analysis}
\section*{Introduction}\label{sec1}

Nanofluidic devices with palladium islands combine asymmetry and confinement, resulting in distinctive physics, reflected in the very high rectification and abrupt threshold. Specifically, under sinusoidal excitation, they display a strongly nonlinear behavior, with a charge threshold followed by a rapid shift to a highly conductive state. We observed experimentally that this threshold is explained by mechanical deformations of the device, in which a liquid-filled blister reversibly forms. This blister responds to applied voltages and causes large conductance changes, including the abrupt threshold observed. We attempt here to rationalize (1) the device behavior in the OFF state, and (2)  the ON/OFF conductance modulation through the blister dynamics. Then, we provide a plausible mechanism for the blister formation which involves (3) excess cation focusing to produce large net charge densities and (4) Coulomb repulsion overcoming the graphite/palladium adhesion.\\

\subsection{The OFF-state conductance}\label{secOFF}

We model the device in the OFF-state (in the absence of blister) as a hollow cylindrical slab of inner radius $r_\text{in}\sim 50$ nm, outer radius $r_\text{out}\sim25$ $\mu$m and height $h\approx5$ nm (Supplementary Figure 24), following typical device dimensions extracted from TEM images as shown in Supplementary Figures 2-3. We define the aspect ratio of the device as $a=r_\text{out}/r_\text{in}=500$. The channel entrances exhibit a non-zero surface charge $\Sigma=100 $ mC/m\textsuperscript{2} in accordance with previous reports for concentration-regulated surface charge on graphite\cite{secchi2016scaling,emmerich2022enhanced}. A bias is applied to the device such that the electrostatic potential's value at the inner entrance is $\Delta V$ and its value at the outer entrance is 0.

\begin{figure}[h]
    \centering
    \includegraphics{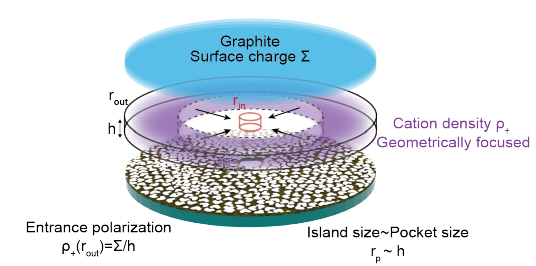}
    \caption{Key ingredients of the OFF-state model. The top graphite wall is charged and the entrance surface charge regulation ensures a net cation excess which is geometrically focused under the application of an inward electric field.}
    \label{fig:my_label}
\end{figure}

We first compute the OFF state conductance of the device, before moving to the mechanism for an increased conductance. We obtain this conductance through writing the following equations:
\begin{enumerate}
    \item $E(r)=-\nabla V(r)$
    \item $j(r)=\sigma E(r)$
    \item $\nabla \cdot j(r)=0=\frac{1}{r} \,\partial_r  \big(rj(r)\big)$
\end{enumerate}
Where $E(r), V(r), j(r)$ are the electric field, the potential and current density, respectively. $\sigma$ is the bulk conductivity of the electrolyte. For KCl at 1M, $\sigma =10S/m$. 

We have $I=2\pi rhj(r)=\text{constant}$, therefore $E(r)=I/(2\pi rh\sigma)$. The potential thus reads $V(r)=\int_r^{r_\text{out}} E(r) dr =I/(2\pi\sigma h) \ln(r/r_\text{out})$. Using the boundary condition  $V(r_\text{in})=\Delta V$,  we obtain the value of the device conductance:
\begin{equation}
G_\text{OFF}=I/\Delta V=2\pi \sigma h / \ln a
\end{equation}
Where $\ln(a)\approx 4.6$.  As the conductance of a nanochannel with height $h$, width $w$ and length $L$  is $G_\text{channel}=wh/L \times \sigma$, we can note that the Pd island device conductance is equivalent to that of a nanochannel with dimensions $h$, $w=2\pi r_\text{in}$ , $L=r_\text{in} \ln a \approx 500$ nm. That is, \textbf{the conductance level of the device is governed by the inner entrance region}.\\

Evaluating this expression for reasonable parameters gives a conductance level close to the experimental OFF state. Therefore, the over-conducting ON is obtained with a modified geometry effectively leading to a partial short-circuit of the channel considered above.\\

Moreover, we obtained the full spatial distribution of the electric field:
\begin{equation}
    E(r)=\frac{\Delta V}{r \ln a}
\end{equation}
We observe that the electric field developed close to the inner wall is $\Delta V/(r_\text{in} \ln a)$ which is much larger than the field in a nanochannel with length $r_\text{out}$, by a field enhancement factor $a/\ln a \approx 80$.\\

\subsection{ON/OFF conductance modulations and threshold due to blister displacement}

The blisters observed experimentally give rise to thin film interference patterns with several fringes, indicating that the optical path difference between the light reflected above (SiN window) and below the blister (2D crystal) is on the order of a few wavelengths. Therefore, we observe micron-scale deformations. Ionic transport within the dynamical micro-channel formed by the blister is therefore unconfined and can be considered as a zero-resistance path. In other words, the blister short-circuits the nanofluidic circuit formed by the nanochannel network in between Pd islands. 
Depending on where the blister forms, it can have different effects on the device conductance. If it forms close to the pore, it will reduce the pore entrance and rescale the off-state conductance (as shown in section \ref{secOFF}).   Although a memristive response is expected in this picture, it does not explain the threshold as the blister can form gradually around the pore region. When the blister forms off-center, pronounced thresholds were observed and correlated experimentally with the blister crossing the pore region (threshold scenario 1, Supplementary Videos 1 and 2). We have observed another scenario leading to a threshold uncorrelated with the pore region crossing (scenario 2), which we attribute to a blister extension outside of the window region (threshold scenario 2, Supplementary Videos 3 and 4)

For the devices exhibiting OFF-state conductance values compatible with the above description, we observed that the shift to the ON-state conductance matches the passage of the blister over the pore entrance region (threshold scenario 1). We can then rationalize the ON-state conductance in a simplified geometry as proposed in Figure 3d. When a small, off-centered blister of size negligible compared with the window size is present, it hardly affects the device conductance, and the conductance is very close to its OFF-state value $G_\text{OFF}=2\pi \sigma h / \ln{(r_\text{out}/ r_\text{in})} $.  When a blister with radius $r_\text{blister}$ forms, it forms a short circuit such that the conductance is given by the remaining adhered channel network outside the blister, which reads:
\begin{equation}
    G_\text{ON} = 2\pi \sigma h / \ln{(r_\text{out}/ r_\text{blister})}
\end{equation}
We can therefore note that the loss of resistance is simply given by the resistance of the formed blister: $R_\text{OFF}-R_\text{ON}=\ln{(r_\text{blister}/ r_\text{in})}/ 2\pi \sigma h $
In this picture, the conductance ratio is given by:
\begin{equation}
   G_\text{ON}/G_\text{OFF}=\frac{\ln{(r_\text{out}/ r_\text{blister})}}{\ln{(r_\text{out}/ r_\text{in})}}
\end{equation}
This mechanism explains both the ON and OFF state, and the steepness of the switching threshold between these states is specific to the blister dynamics and position, which may vary from device to device. After these phenomenological considerations driven by \textit{in-operando} imaging, we now rationalize the blister formation. 

\subsection{Blister formation: focusing charges to overcome adhesion}
We propose that the observed blisters arise from electrostatic repulsion overcoming the adhesion force between the palladium islands and the 2D crystal. In the following sections, we show analytically that in the presence of a strong surface charge and confinement, it is possible to focus excess counter-ions geometrically. Then, we demonstrate that the obtained charge densities may be large enough to overcome adhesion. 

\subsubsection{Charge focusing enabled by the radially converging geometry}

We show that the entrance counter-ions balancing the outer entrance surface charge can be focused geometrically by radially converging electric fields to higher concentrations inside the channels, which may lead to the breakdown of electroneutrality in the channels. This is reminiscent of the so-called 'spatially charged zone'  asymmetric nanocapillaries\cite{jubin2018dramatic}. The non-zero (negative) surface charge $-\Sigma$  of outer walls brings the local electroneutrality balance at $r=r_\text{out}$:
\begin{equation}
    -\Sigma + \rho_{+,D}^o \lambda_D = 0
\end{equation}
Which means there is a non-zero (positive) counter-ion density $\rho_{+,D}^o=Fc^o_+$  within the Debye layer close to the outer walls.
Upon drifting towards the device center, we assume that  ions will no longer be located within the Debye layer of thickness $\lambda_D$ close to the wall, but rather dispersed over the whole height of the channel, leading to an overall decrease of the height-averaged concentration $\rho^o_+=\rho_{+,D}^o \times \lambda_D/h$. In other words, for simplicity, we neglect out-of-plane concentration variations in the device driven away from electroneutrality, and simply write the boundary condition:
\begin{equation}
\label{eqBoundaryConditions}
    \rho_+(r_\text{out}) = \Sigma/h = \rho^o_+
\end{equation}
We assumed here that at the outer entrance of the device, the diffuse counter-ion layer reaches equilibrium very quickly as it is in contact with the solution, ensuring the boundary condition equation \ref{eqBoundaryConditions} 
 is verified at any time. However, the excess charge within the device is out of surface charge regulation equilibrium as it is not in contact with the reservoir.
The exact fate of the excess charge distribution $\rho_+(r,t)$ under sinusoidal voltage cannot be easily computed, but we derive the result for the DC bias case below, showing that this charge can indeed be focused to higher densities geometrically.

\subsection{Radially converging drift-diffusion DC model}

\subsubsection{Drift-diffusion in cylindrical coordinates}
The full drift-diffusion equation in cylindrical coordinates reads:
\begin{equation}
    \partial_t \rho_+ + v(r) \partial_r \rho_+ = D\big( \partial_{rr} \rho_+ + \frac{1}{r}\partial_r \rho_+ \big)
\end{equation}
Treating the excess charge density as a perturbation without back-action on the electric field distribution, the charge velocity field reads: $v(r)=\mu E(r) = \mu \Delta V/(r \ln a):=U /r$. The constant $U=\mu \Delta V/\ln a$ is negative for a positive bias applied, as we have inward motion of the excess cations.

\subsubsection{Analytical solution}

Cahn and Jackson\cite{cahn2008solvable} obtained an analytical expression for an ion distribution undergoing inwards radial drift-diffusion. The analytical solution of the problem for a $\delta$ point source reads:

\begin{equation}
    \rho^0_+(\vec{r},s)= \frac{1}{\sigma_\parallel(s) \sqrt{2\pi}} e^\frac{-\big(r\cos{\theta}-r_o(s)\big)^2}{2\sigma^2_\parallel(s)}\times \frac{1}{\sigma_\perp(s) \sqrt{2\pi}} e^\frac{-r^2\sin^2{\theta}}{2\sigma^2_\perp(s)}
\end{equation}
Where $\vec{r}\equiv (r,\theta)$ in polar coordinates, $s=t/T$, $T$ is the electrophoretic time $T=R_\text{out}^2 \ln \alpha/2 \mu \Delta V$,   $r_o(s)=r_\text{out}\sqrt{1-s}$, $\sigma_\parallel=\sqrt{\frac{2 D T s}{1+g_\parallel(s)}}$, $\sigma_\perp=\sqrt{\frac{2 D T s}{1+g_\perp(s)}}$, $g_\parallel(s)=\frac{1+s}{2+s}$, $g_\perp(s)=-1+\frac{s}{(1+s)\ln{(1+s)}}$, as derived by Cahn and Jackson\cite{cahn2008solvable}.

\begin{figure}[h]
    \centering
    \includegraphics[width=\textwidth]{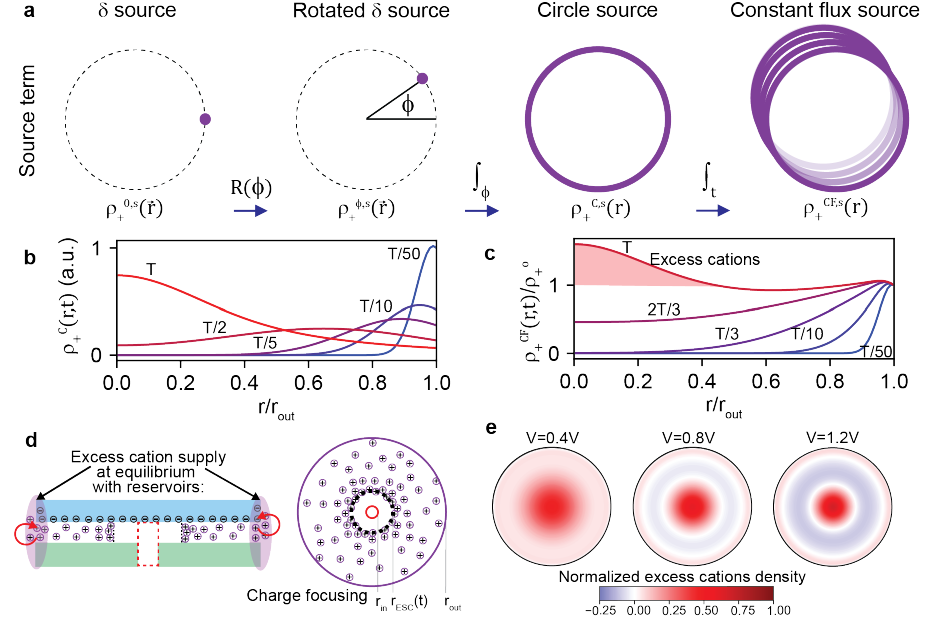}
    \caption{\textbf{DC model for geometry-induced cation focusing in HACs.} \textbf{a}, Construction of the
constant flux source by superposition of $\delta$ sources.\textbf{ b}, Time evolution of the circle source 
at t=T/2 and t=T (DC breakthrough time). \textbf{c}, Time evolution of the constant flux source, with the red highlighted part at t=T illustrating the excess cations exceeding the surface charge equilibrium. \textbf{d}, Sketch of the
charge focusing, with a source at the outer entrance supplied by surface charge regulation at equilibrium with reservoirs. \textbf{e}, Radial profile of the excess charge charge distribution at t=T for different voltages V=0.4, 0.8 and 1.2V, normalized by the outer wall surface charge.}
    \label{fig:SImodelfocusing}
\end{figure}

More generally, the solution for a $\delta$ source at time 0 and position $(r_\text{out} \cos \phi,r_\text{out} \sin \phi)$ in Cartesian coordinates reads:
\begin{equation}
    \rho^\phi_+(\vec{r},s)= \frac{1}{\sigma_\parallel(s) \sqrt{2\pi}} e^\frac{-\big(r\cos{(\theta-\phi)}-r_o(s)\big)^2}{2\sigma^2_\parallel(s)}\times \frac{1}{\sigma_\perp(s) \sqrt{2\pi}} e^\frac{-r^2\sin^2{(\theta-\phi)}}{2\sigma^2_\perp(s)}
\end{equation}
As sketched in Supplementary Figure 25.a, we obtain the circle source solution by integrating over the angle $\phi$:
\begin{equation}
    \rho^C_+(r,t)\sim \int_0^{2 \pi} \rho^\phi_+(\vec{r},t) d\phi
\end{equation}
Where the integration over $\phi$ suppressed the dependency over $\theta$, granting the distribution radial symmetry.
To obtain the solution with a constant flux of ions, this solution can be integrated with respect to time, yielding the constant flux source solution:
\begin{equation}
    \rho^{CF}_+(r,t)\sim \int_0^t \rho^C_+(r,t') dt'
\end{equation}
We evaluated $\rho^C_+$ and $\rho^{CF}_+$ numerically, to check whether the converging geometry enables charge focusing.
Our analytical results demonstrate the possibility of geometrically focusing the outer entrance charge with the radially converging geometry, in such a way that the center region has an unbalanced charge density $\rho^{CF}_+ - \rho_+^o$ where $\rho_+^o=\rho_+(r_\text{out})=\Sigma/h$ is the part of the net charge that counters the wall surface charge $\Sigma$. As shown in  Figure \ref{fig:SImodelfocusing}c and \ref{fig:SImodelfocusing}e, we obtain charge densities in the central region well above the outer wall surface charge, which open the door for unscreened Coulomb repulsion in the channels.


\subsubsection{Blister formation at large charge densities}

We demonstrated that we can up-concentrate counter ions in highly asymmetric channels with a radial geometry. We propose that the experimentally observed mechanical deformations arise from an out-of-equilibrium net charge density in the channels induces a \textbf{electrostatic pressure} which can overcome the van der Waals adhesion between the graphite and the Pd islands. The diversity of observations is largely modulated by the silicon nitride membrane strain distribution, which can provide yield points different that the center of the radially converging flow of counter-ions.

Inside a spherical pocket with radius $r_p=5$ nm (size between the Pd islands) we have an excess counter-ion concentration $c_+$ and a net charge density $\rho_+=Fc_+$ where F is the Faraday constant. Summing Coulomb repulsion energies over all ions within the pocket, we obtain a total energy of:
\begin{equation}
    E_+= K \frac{\rho_+^2\ r_p^5}{\epsilon_o\ \epsilon_r}
    \end{equation}

 where the prefactor $K$ is equal to $2\pi/15$ in a spherical pocket geometry. The volumetric energy corresponds roughly to an equivalent pressure of $P_+ \sim K \frac{\rho_+^2\ r_p^2}{\epsilon_o\ \epsilon_r}$. Metal islands are expected to screen any longer-range interaction between charges, making the interstitial pocket size the relevant dimension here. 
 To get an order of magnitude of the magnitude of the effect, we can examine the expected pressures that should build up in the system with counter-ions concentrations accessible by the system. 
We expect an outer entrance outer surface charge $\Sigma=100$ mC/m\textsuperscript{2} for graphite in 1M KCl (value extrapolated from Emmerich et al\cite{emmerich2022enhanced}). This translates into a concentration $c^o_+=\Sigma/F r_p \approx 300$ mM.

We write the electrostatic pressure as a function of the concentration of unscreened cations $P_+ \sim K \frac{F^2 r_p^2}{\epsilon_o \epsilon_r} c_+^2$ and derive the proportionality coefficient between the pressure in bars and the counter-ion concentration in mM. This coefficient is equal to ${10}^{-5}\times{10}^6\times K\frac{F^2r_p^2}{\epsilon_o\epsilon_r}\sim 1$ mbar/mM\textsuperscript{2}. Therefore, while modest concentrations of \~10 mM cause little electrostatic pressure (100 mbar), the quadratic dependency on the cation concentration can bring large pressure increases with an expected value of 10 MPa for 300 mM cation concentration. This is comparable with pressures estimated in 2D materials blisters \cite{sanchez2018mechanics}.
We consider the appearance of a small blister of surface $dS$, small enough to consider that $c_+$ and thus $P_+$ are uniform, but large enough to consider that once the blister is formed, the excess charge is highly diluted and thus negligible. This should correspond to the 1-micrometer scale. We consider the energy change which reads: 
\begin{equation}
    dE=+\Gamma\times dS- K \frac{ \rho_+^2  r_p^2}{\epsilon_o\ \epsilon_r } \times r_p \, dS
\end{equation}
where the first surface term corresponds to the adhesion energy increase and the second volumetric term corresponds to the Coulombic energy decrease.

The blister formation becomes favorable when $dE/dS<0$, that is the system can reduce its energy by forming the blister. The concentration threshold to obtain the blister formation reads:
\begin{equation}
\rho_+^\ast=\sqrt{\frac{\Gamma\epsilon_o\ \epsilon_r}{K r_p^3}}
\end{equation}
Using an adhesion energy $\Gamma=20$ mJ/m\textsuperscript{2} in the range of 10\% of measured (dry) graphite-metal adhesion energies\cite{megra2019adhesion}, we obtain a critical blister formation concentration in the range of 0.5M, which should be attainable by a 2-fold concentration of the surface charge due to the radially converging flow. 

Merging the radial charge focusing picture with this instability reasoning for blister formation, we can group the adhesion and surface considerations as follows.
The maximum charge density that builds up in the device is $\rho_+^\text{max}=\Phi \rho_+^o = \Phi \Sigma /h$. Therefore, the criterion $\rho_+^\text{max}>\rho_+^\ast$ for blister formation becomes:
\begin{equation}
    \Phi \Sigma /h >\sqrt{\frac{\Gamma\epsilon_o\ \epsilon_r}{K r_p^3}}
\end{equation}
We assimilate the device height $h$ and the pocket size $r_p$ and obtain the following dimensionless criterion:
\begin{equation}
\label{eqBlisterFormation}
    K \Phi^2 \times \Sigma^2 h / \Gamma \epsilon_o \epsilon_r > 1
\end{equation}
Therefore, the device behavior is essentially governed by the ratio $\Sigma^2/\Gamma$, which should be large enough to enable the blister formation. This seems to be the case for graphite devices but not for mica ones. The other parameters depend on the device geometry and channel network properties ($K, \Phi, h)$. While we assumed for simplicity that the dielectric constant of water was its bulk value $\epsilon_r=80$, it may be reduced for the case of interfacial water\cite{fumagalli2018anomalously}, which would further enhance the predicted effects. \\

These predictions are in line with the experimental observation that large KCl concentration were required to obtain the memristive behavior of the devices (Supplementary Figure 14), as KCl is expected to regulate the surface charge of the 2D crystals. For time-dependent AC measurements, we rationalize the charge threshold reported in Figure 2 as follows: the blister formation criterion eq. (\ref{eqBlisterFormation}) occurs fast at positive voltages and the charge threshold  relates mostly to the blister dynamics where enough charges need to be brought to enlarge the blister past the pore. The threshold corresponds to flushing all ions and a possible delay for the blister dynamics, both of which are governed by the amount of excess charge focused into the device.  \\

\textbf{Where the blister forms}\\
While the maximum concentration is expected to happen at the narrowest constriction - the pore entrance – we do not systematically observe that the blister forms precisely there. This observation can be rationalized as the radially convergent excess counter-ion up-concentration mechanism allows for parts of the device to reach the concentration threshold before the cation front reaches the pore. Moreover, as the actual geometry is not perfectly radial, the focusing might happen at pre-determined positions. An additional effect not taken into account so far is the possibility for the silicon nitride membrane to be pre-strained due to growth\cite{temple1998residual}. The pre-strain energy could assist in peeling the blister, thereby locally lowering the concentration threshold for blistering derived above. This pre-strain could not be controlled in our experiments and may explain variations in our devices regarding the location of the blistering and the abruptness of the threshold. However, it represents an avenue for further optimization of the devices in a controlled manner. \\

\textbf{Second threshold mechanism: }
For devices presented in Supplementary movies 3 and 4, the threshold does not correlate with the blister passing over the pore, but rather with an abrupt volumetric change of the blister. The blister edge displacement accompanying the threshold is then located close the window edge. In this scenario, we attribute the conductance change to a blister extension close to the top flake edge rather than close to the pore center. This second mechanism yielding similar IV characteristics (with however a higher $G_\text{OFF}$ as the blister is over the conductance-limiting region close to the pore) points to the importance of \textit{in-operando} observations, and that of optimizing devices to control the blister motion
which governs the memristor performance. 
\pagebreak

\section{Ionic logic}
\subsection{Setup}
The details of the memristive logic circuit for the IMP gate implementation are shown in supplementary Figure 26. A source measurement unit (SMU, Keithley 2636B) provides the voltage imputs $V_\text{q}$ and $V_\text{p}$ for the two Q and P memristors involved in the IMP operation. In particular, all terminals belonging to the ionic cells consist of chlorinated tips connected by metal clips to the rest of the circuit. The metal clips are connected by wires to a printet circuit board (PCB) with header pins and BNC connectors. The BNCs are connected to the SMU, providing the Q and P terminals, as well as the ground. The common terminal shared between P and Q is connected to a variable electronic resistor. The other terminal of the variable resistor is connected to the instrument ground. The SMU allows us to force $V_\text{q}$ and $V_\text{p}$ while simultaneously reading $I_\text{q}$ and $I_\text{p}$, thus retrieving the data presented in Figure 4 of the main text.

\begin{figure}[!h]
	\centering
	\includegraphics[width=1\linewidth]{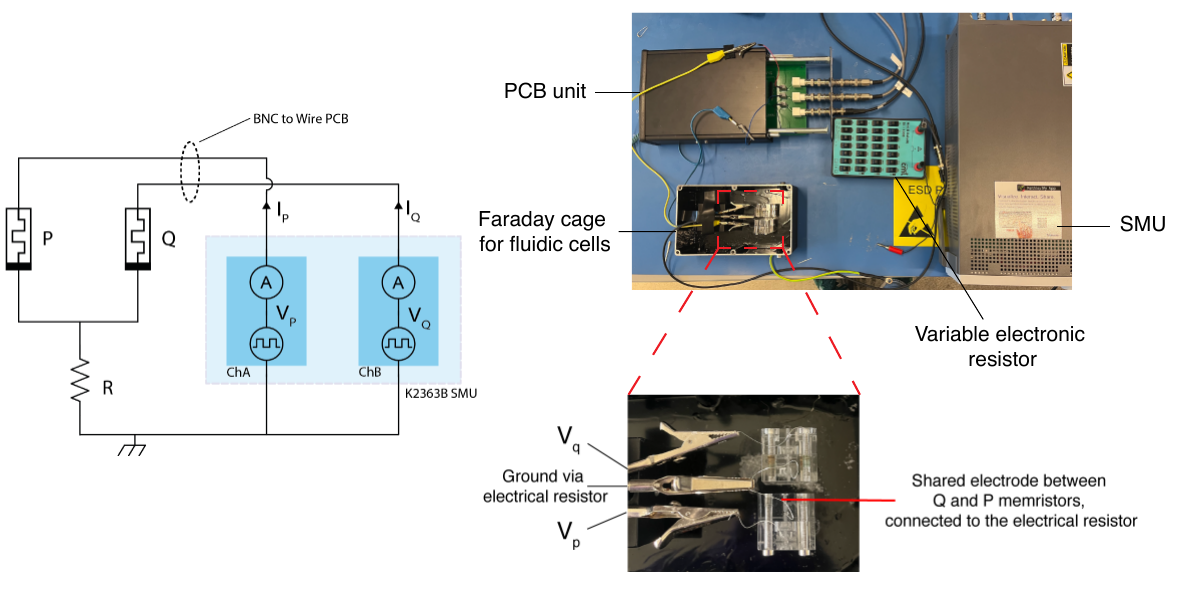}
	\caption{ \footnotesize	\textbf{Setup for ionic logic. Left:} Details schematic of the circuit. P and Q represent both HAC channels. The electronic resistor is drawn in black. \textbf{Right:} Photography of the complete setup with a zoom on fluidic cells below.}
\end{figure}

\pagebreak
\subsection{One HAC with a resistor in series}

In the logic circuit drawn in Figure 4.a, HACs are connected in series with a variable resistor set to a few megaohms. As stated in the main text, a resistor placed in series with a memristive device will have a lower conductance ratio than the memristor alone. Upon the application of voltage, the memristor's conductance and thus the current will increase. This has the effect of increasing the voltage drop across the resistor hence decreasing the one on the memristor. We therefore connected a HAC in series with a variable electronic resistor in order to verify that the combination of these two component still behave as a memristor (Supplementary Figure 27.a). We used the same equipment as for the logic circuit (presented in the previous section) but plugged only one memristor. We find that HACs in series are still able to operate in this configuration, yet at the costs of a decrease in conductance ratio, which enables performing the IMP logic gate (Supplementary Figure 27.b).

\begin{figure}[!h]
	\centering
	\includegraphics[width=1\linewidth]{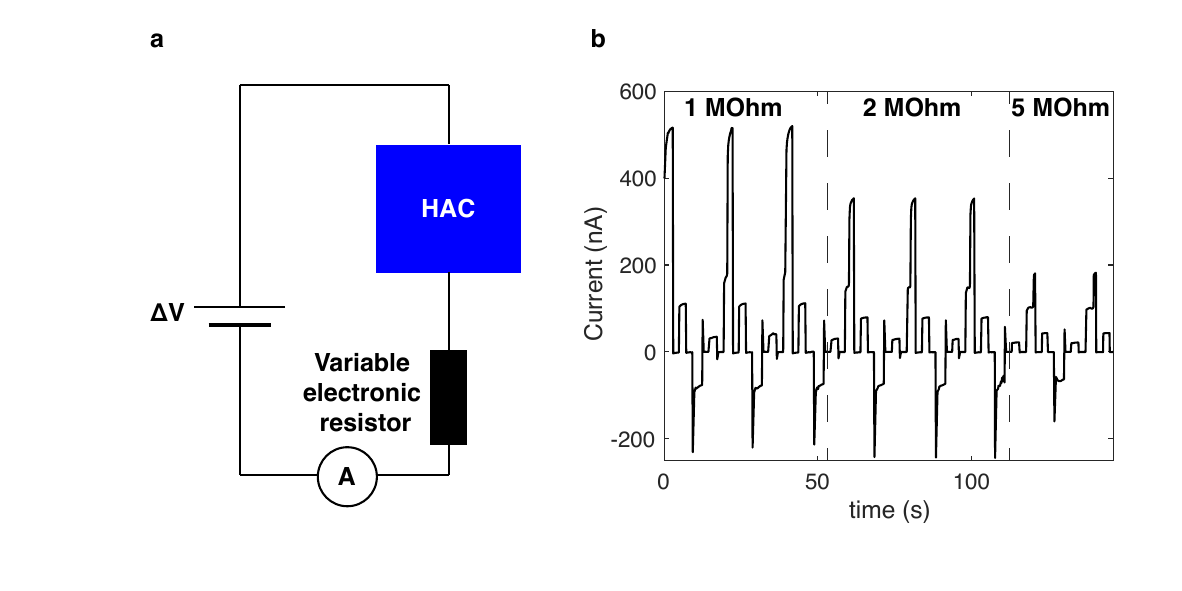}
	\caption{ \footnotesize	\textbf{HAC with a resistor in serie (Device 29-1M KCl). a,} Schematic of the circuit. \textbf{b,} Pulse-programing performances for different settings of the variable electronic resistor. Voltage pulse programming similar as in supplementary Figure 6 are applied repeatedly. The value of the electronic resistor is manually changed during the experiment.}
\end{figure}
\pagebreak

\subsection{Additional data with two HACs}

For Robustness regarding ionic logic, we repeated the experiment presented in main text Figure.4, focusing only on the non-trivial cases.  The results are presented in Supplementary Figure 28. For this pair of the device, the charge threshold was of lower value than the ones presented in the main text. This imposed working with pulses of lower amplitude and duration. In this experiment, the resistor was set to 4 MOhm.

\begin{figure}[!h]
	\centering
	\includegraphics[width=1\linewidth]{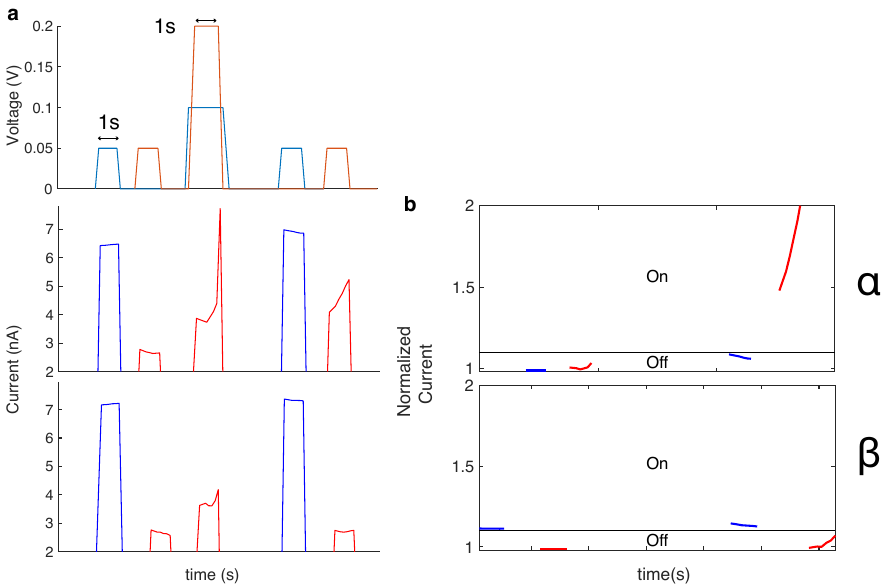}
	\caption{ \footnotesize	\textbf{Supplementary data for ionic logic with two HACs (Devices 30 and 31-1M KCl). a,} Up: Applied potential for the P-memristor (in blue) and Q-memristor (in red). Middle: raw current data for the $\alpha$ case (P=0, Q switch). Down: Raw current data for the $\beta$ case (P=1, Q does not switch).   \textbf{c,} Normalized current data for the two cases presented in \textbf{b}. Each memristor's current pulses is normalized by its minimum value. The limit between ON and OFF states is placed at 10$\%$ of relative variation.}
\end{figure}
\pagebreak

\subsection{Additional data with one HAC and one tunable electronic resistor}
For robustness regarding ionic logic, we also performed experiments with one HAC and two variable electronic resistors, as shown is supplementary Figure 29.a. The first electronic resistor plays the role of the P-switch and its conductance is manually changed during the experiment depending on the targeted state. Here we used 1 MOhm as ON state conductance and 10 Mohm as OFF state conductance which matches HACs. The second electronic resistor is set to 2 MOhm and remains fixed as for experiments with two HACs. One HAC is used as Q-switch. We find that we are indeed able to perform both non-trivial cases of IMP logic (supplementary Figure 29.b).

\begin{figure}[!h]
	\centering
	\includegraphics[width=1\linewidth]{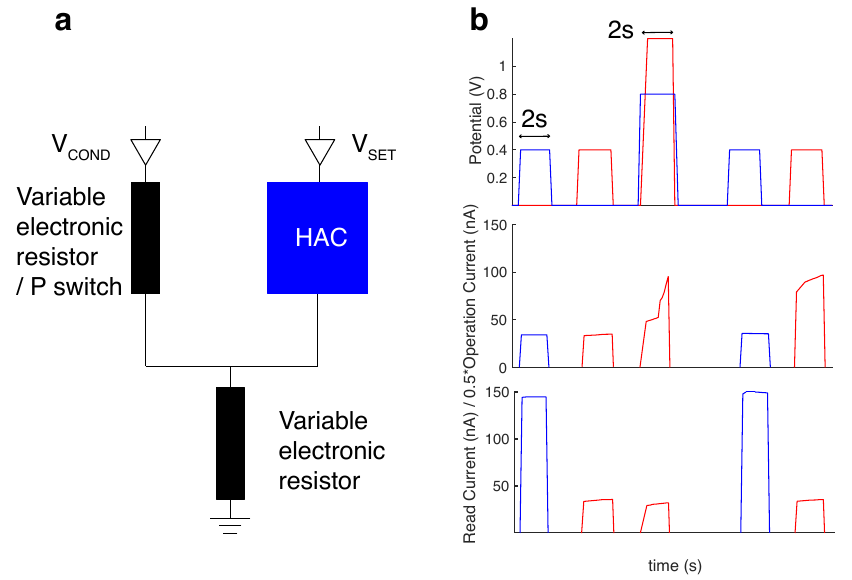}
	\caption{ \footnotesize	\textbf{Supplementary data for ionic logic with one HAC as the Q-switch and one programmable resistor as the P-switch (Device 32-1M KCl). a,} Schematic of the circuit. \textbf{b,} Raw data for IMP logic with applied potentials (up) and measured current for $\alpha$ case (middle) and $\beta$ case (down).}
\end{figure}

\subsection{Numerical simulation}
In this section, we reproduce the ‘stateful’ logic operation with a numerical simulation method under a similar condition that we have with ionic memristors. This simulation is based on the \textit{Matlab, Simulink}, and \textit{Simscape} packages.
\subsubsection{Numerical model of the switching-threshold memristive device.}

A classical numerical model of memristor in \textit{Simscape} package is an ideal memristor with a nonlinear dopant drift approach\cite{Wolfmemristor}. Here we modify the model to achieve a charge-threshold dynamics. The state of the device will switch only when the total charge is beyond threshold value. The mathematic expression of charge-threshold memristive element can be written as following equations:

\begin{equation}
V=I \cdot M\\
\end{equation}
\begin{equation}
M=\epsilon \cdot R_A+(1-\epsilon) \cdot R_B\\
\end{equation}
\begin{equation}
q=\int I dt\\
\end{equation}
\begin{equation}
\dfrac{d\epsilon}{dt}=
\begin{cases}
0,& \text{if } q<Q\\
\frac{I}{Q_s}F_p(\epsilon) & \text{if } q\ge Q
\end{cases}\\
\end{equation}
\begin{equation}
F_p(\epsilon)=1-(2\epsilon-1)^{2p}
\end{equation}

Where, V is the voltage across the memristor, M is the dynamic resistor value of the memristive element, I is the current go through this device, $R_A$ is the high conductive state resistance, $R_B$ is the low conductive state resistance, $\epsilon$ is the fraction of the memristor in state A, t is the time, $q$ is the total charge that flux through the devices, $Q$ is the charge-threshold to be overcome for initializing the state transition, $Q_s$ is the total charge required to make the memristor transition from one state to another (when the state transition starts), $F_p (\epsilon)$ is a ‘window’ function, which keeps $\epsilon$ between 1 and 0, and realize the zero drift at the boundaries of the device, and p is positive integer to control the exponential scale of state switch.\\

Compared with our modified model with ideal memristor accessible in \textit{Simscape} Package, the devices conduct the stateful transition with the exact tendency but delayed with a consistence value of charge threshold, thus there are only limited influence on the reliability of the model. To be noticed that the definition of $Q_s$ has been modified. Now in our modified model, $Q_s$ and $F_p (\epsilon)$ are used to controlled the shape of the transition curve, whereas the total charged required to the memristor transition from being fully in one state to being fully in the other state can be expressed as $Q_s+Q$.

\subsubsection{Numerical simulation of ‘stateful’ logic operation circuit.}
Based on this charge-threshold memristive model, we connect two of these devices in the following simulated circuit built in \textit{Simulink}. With this simulated circuit, we can direct input the same voltage documents used in experimental section of ‘stateful’ logic operation (Figure 4) to run the simulation. By applying similar performance of real nanofluidic memristors that we have in experimental section (supplementary table 2), we could reproduce a similar circuit with a numerical model (supplementary Figure 30.a). Here, the threshold is several tens of nC and the transition ratio is 10.
Different from real experimental situation, in which devices need to be set by prior voltage pulses, the conductance state can be manually set by defining the value of initial State A fraction $\epsilon_0$, which ‘0’ corresponding to low conductance state and ‘1’ corresponding to the high conductance state. Here the fraction of State A $\epsilon_(t_1)$ at alternative time $t_1$ is defined as following in the simulation:

\begin{equation}
\epsilon_{t_{1}}=\epsilon_0+\int_0^{t_1} \dfrac{d\epsilon}{dt} dt
\end{equation}

Compared with Figure 4, Supplementary Figure 30 reproduce its result within a numerical simulated system. This result could support that, from the objective of simulation, the ‘stateful’ logic operation could be achieve with memristive switches displaying a conductance ratio of 5.

\begin{figure}[!h]
	\centering
	\includegraphics[width=1\linewidth]{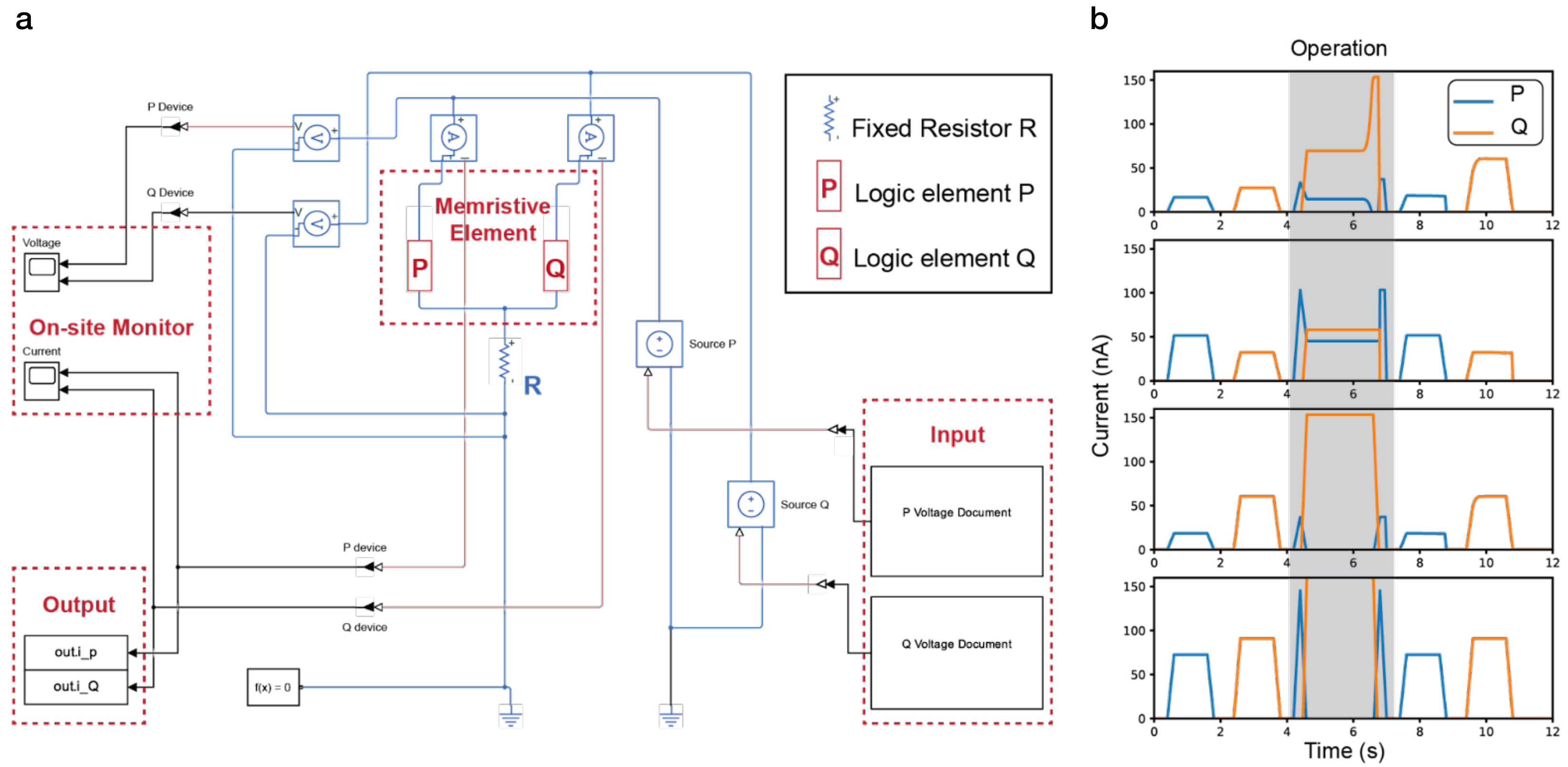}
	\caption{ \footnotesize	\textbf{The ‘stateful’ logic operation circuit model and simulated current trace. a,} Artificial circuit model to simulate ‘stateful’ logic operation, including input section that capable to directly run voltage documents in Figure 4, on-site monitor, and output for current analysis. \textbf{b,} Simulated current of logic operations.}
\end{figure}

\scriptsize
\begin{flushleft}
\begin{tabular}{| m{2cm}| m{2cm} | m{2cm}| m{2cm} | m{2cm}| m{2cm}|} 
  \hline
 Component P & Values & Component Q & Values  & Component R & Value  \\ 
 \hline
  R$_A$ & 4 MOhm &   R$_A$  & 2 MOhm  &  Resistance, R & 6 MOhm  \\ 
 \hline
  R$_B$ & 20 MOhm &   R$_B$  & 10 MOhm  &  & \\ 
 \hline
Q$_s$ & 20 nC &  Q$_s$ & 20 nC  &  &   \\ 
 \hline
 Q & 130 nC & Q  & 130 nC  &  &   \\ 
 \hline
 P & 2 & P & P  &  &\\ 
 \hline
 
\end{tabular}
\normalsize

 \footnotesize{Supplementary table 2: }\textbf{Detailed parameter of resistor and memristive elements in numerical simulated circuit model (Supplementary Figure 29).}

\end{flushleft}

\pagebreak

\bibliographystyle{unsrt}